\documentclass[final,leqno]{siamltex}
\usepackage{amsmath}
\usepackage{svg}
\usepackage{graphicx}

\usepackage{caption}
\usepackage{subfig}
\usepackage{pdflscape}

\usepackage{mathtools}
\DeclarePairedDelimiter\floor{\lfloor}{\rfloor}

\usepackage{algorithm}
\usepackage{algpseudocode}
\usepackage{soul}
\usepackage{color}

\usepackage{varioref}

\newcommand{\hlgray}[1]{{\sethlcolor{lightgray}\hl{#1}}}

\title{Evaluating a sublinear-time algorithm \\ for the minimum spanning tree weight problem}

\author{Gabriele Santi\thanks{Master's Degree in Computer Science, University of Rome ``Tor Vergata'' ({\tt gabriele.santi@acm.org}).}
        \and Leonardo De Laurentiis\thanks{Master's Degree in Computer Science, University of Rome ``Tor Vergata'' ({\tt ldelaurentiis@acm.org}).}}

\begin{document}

\maketitle

\begin{abstract}
We present an implementation and an experimental evaluation of an algorithm that, given a connected graph G (represented by adjacency lists), estimates in sublinear time, with a relative error $e$, the Minimum Spanning Tree Weight of G (see~\cite{crt} for a theoretical exposure of the algorithm). 
Since the theoretical performances have already been shown and demonstrated in the above-mentioned paper of Chazelle et al. our goal is, exclusively, to experimental evaluate the algorithm and at last to present the results.
Some technical insights are given on the implementation of the algorithm and on the dataset used in the test phase, hence to show how the experiment has been carried out even for reproducibility purposes;
the results are then evaluated empirically and widely discussed, comparing these with the performances of the Prim algorithm and the Kruskal algorithm, launching several runs on a heterogeneous set of graphs and different theoretical models for them.

We assume hereafter that the reader has knowledge about the cited paper as we will just recap the theoretical results.
\end{abstract}



\pagestyle{myheadings}
\thispagestyle{plain}
\markboth{\scriptsize GABRIELE SANTI AND LEONARDO DE LAURENTIIS}{\scriptsize EVALUATING A SUBLINEAR-TIME ALGORITHM FOR THE MST WEIGHT PROBLEM}

\section{Introduction}
We will discuss here some preliminary observations and assumptions. First of all, we observe that we need a set of graphs that satisfies the following points:
\begin{itemize}
 \item they should be finite and should not be multigraphs;
 \item they should be undirected;
 \item they should have weighted edges;
 \item the weights on the edges should be integers; since the graph is finite, it is enough to show that there exist $W$ such that it is the maximum weight on the edges of the graph;\footnote{more generally, it is enough to have a numerable set of values for the weights which is always true when the graph is finite}
 \item they might contain self-loops;
 \item they have to be connected (the graph has only one connected component);
 \item they should be represented with adjacency lists;
 \item they should be represented in the same manner (we need an unvarying file format).
\end{itemize}
Unfortunately, the graphs and their representations, which can easily be found on the Internet, don't accomplish all of this requirements at the same time, although many standards for the file format are available.

Given this observation, our choice was to use randomly generated graphs, hence to implement our own graphs generator. This gives us the opportunity to generate a wide set of connected graphs, with tunable parameters, carefully chosen looking forward to the tests; these parameters include the number of nodes, the number of edges and the edges weight, nonetheless the distribution law for the edges. The edges between the nodes are step-by-step randomly constructed, respecting the connection requirement. The different types of graphs that we use in our experimental evaluation are presented afterwards.

After studying the paper we made some assumptions. One of the problem we encountered is that the theoretical algorithm assumes to have as input only graph $G$ and to have direct access to the family of graphs $G_i$\footnote{we recall that $G_i$ is an induced subgraph of $G$ such that the maximum weight on his edges is $i$ (e.g. $G_w$ is exactly $G$)}; with ``direct'' we intend that no computation is required to extract $G_i$, which is not true. In fact, we can show easily that a lower bound for the extraction of all the family is, at least, $O(m)$ (i.e. is linear on the number of edges). A na\"{\i}ve approach would cost $O(m)$ for the extraction of $G_i$ for a given $i$, hence $O(w m)$ for the whole family; a better approach could order the edges in $O(m \log m)$ and build the family in a single pass on the edges, achieving $O(m + m \log m)$. Having as input only $G$, it would seem that the algorithm is responsible for the extraction of the family, but this is not desirable due to this lower bound that would sabotage the overall performance. Finally we decided to consider this cost as a part of the construction of the data structure of the graph, to be done prior to the call of the algorithm.

\section{Design studies and choices} 

\subsection{Random Graphs Generator}
As mentioned before, our choice is to implement our own graphs generator. The aim is to generate \emph{connected random graphs} with a specific set of parameters, like the number of nodes, the number of edges, the maximum edges weight and the average degree. Moreover, we want to test how our algorithm behaves in different environments, so we want to control the distribution law of the edges. Keep in mind that the graph has to satisfy all the points of the previous section, among which the connectivity requirement. Given a desired number of nodes $n$ and $e \geq n-1$ edges, the key-concepts under the implementation of our connected graphs generator are the following: 
\begin{itemize}
 \item we begin by generating a random permutation of the vertices $v_0, v_1, \dots v_n$
 \item we generate a random spanning tree by iteratively adding edges in this vector in a uniform random manner; suppose we have added $\tau$ vertices to the tree $T$, we randomly select a vertex $s$ in the set $\lbrace T_0, \dots, T_\tau \rbrace$ and add a new edge $<T_s, v_{\tau+1}>$. At the end we will have an acyclic graph with $n-1$ edges.
 \item following a certain probability law, we add the remaining $e - (n-1)$ edges
\end{itemize}
note that every time we add an edge, a weight is associated to it, with a value uniformly choosen in $\left[ 1, w \right]$.

We even decided to use a custom file format to save the data, which is a \texttt{.ssv} file, standing for \emph{space separated values}: every line of the file corresponds to two edges, reporting the values $<v_s, v_t, w>$ that means source and target nodes, edge weight. Being the graph undirected, it is implicit that the edge $<v_t, v_s, w>$ also exists.

\subsection{Graph Data Structures}
We implemented two versions of the algorithm: the first time using the well-known \emph{Boost}'s BGL\footnote{\texttt{Boost Graph Libray}, \cite{bst}} for the graphs data structures and efficient implementations of Kruskal's and Prim's algorithms; the latter instead embodies our own implementation of both the structures and all the side algorithms. We decided to do so in order to obtain more control over the code for testing purposes and because, at the moment (version \texttt{1.61.0}), the BGL subgraphs framework presents some still unfixed bug.

Unfortunately, our Kruskal algorithm, although using \emph{union-find} structures with path compression\footnote{it is known that the Kruskal algorithm time complexity is $O(m \log m)$, but can be improved to $O(m \log n)$ using union find structures with some euristic conveniently applied}, is not as fine-tuned as that proposed by the Boost's libraries; we didn't take further time on this because on the other side, our version of Prim's algorithm shows the \emph{same exact performances} of the Boost version and it counts as a theoretical lower bound, being an optimal algorithm and way better than the Kruskal's solution. Our data structures have been called FastGraph for their optimization over the operation required for the test. In any way our data structures \emph{can} nor \emph{do} ``help'' the CRT\footnote{Short for B. Chazelle, R. Rubinfeld, and L. Trevisan.\cite{crt}} algorithm in improving his time complexity.

\subsection{Tuning the algorithm}
In the implementation of the algorithm, the graph stored in the text file is read into a \texttt{FastGraph} structure, which is of course a representation by adjacency lists.
We want to compare the performances of CRT algorithm with a standard MST algorithm that in addition computes the total weight.\\
We want to emphasize now that the CRT algorithm is based on probabilistic assumptions; some parameters, that depend asymptotically on $\varepsilon$ should be selected carefully in order to
provide a good approximation very fast. These includes: 
\begin{itemize}
\item $r$, the number of vertices uniformly choosen in the ``approx-number-connected-components''. This number is critical for the performances, as it directly determines the running time of the entire algorithm.
\item $C$, the large costant that we use to pick the number of vertices determinants for the application of the original paper's Lemma 4.
\end{itemize}

Both these values largely affect the overall performances because they undirectly decide how many BFS will be carried out by the CRT algorithm; the BFSes represent the fundamental cost driver of the entire run.

Initially in our opinion, the choice of these parameters must depend on the number of vertices in a way that they are dynamic, keeping their dependency on $\varepsilon$. We tried to untie this bond to $n$ but having static values for this parameters showed poor performances and a relative error exploding too fast as we grow the input instance.
 
As the paper says, the only requirement for $r$ is that it is $O(\frac{1}{\varepsilon^{2}})$ and $C \in O(\frac{1}{\varepsilon})$.

The demonstration reported on the original paper bound these values in a way that helped us on choosing the right function to compute them; in fact, we have that\footnote{see Theorem 7 and Theorem 8 of \cite{crt}}
\[
 r \sqrt{\frac{w}{n}} < \varepsilon < 1/2, \qquad \frac{C}{\sqrt{n}} < \varepsilon < 1/2
\]
hence we choose 

\[
 r = \floor*{ \frac{\floor*{ \sqrt{\frac{n}{w}} \varepsilon - 1 }}{\varepsilon^2}} \in O \left( \frac{1}{\varepsilon^2} \right)
\]
\[
 C = \floor*{ \frac{\floor*{ \sqrt{n} \varepsilon - 1}}{\varepsilon}} \in O \left( \frac{1}{\varepsilon} \right)
\]

\subsection{Random Sequences}
Another problem we encountered was the random selection of $k$ distinct vertices out of $n$ with $k<n$; after some research, we found that this kind of problem cannot be solved in sublinear time on the number of vertices, which can't be done for the same reasons exposed in the introduction about the subgraph family. The problem here is that we can't extract \emph{with no reinsertion} $k$ values without using an external data structure; this structure has a linear cost for the maintainance that depends on $n$. A special case is that in which $n$ is a prime that fulfills certain properties; the key idea is to make use of the properties of the quadratic residues of $n$\footnote{this is widely used in cryptography for ``format preserving encryption''; see~\cite{bbs} and~\cite{cafd}}. That permits us to extract a non-repeating random value in constant time in the range $0 \dots n-1$; sadly, this solution is not affordable here because we need a dynamic range for each call, so that we cannot fulfill such constraints for $n$.

The solution we found is to use Fisher-Yates sequences\footnote{originally decribed in \cite{fys} and then redesigned for computer use in \cite{fysc}} which permits us to prepare in advance the sequences and then get different, distinct values at each call in constant time and with dynamic bounds. The cost of the preparation of those sequences is linear and is not considered in the total cost.

\section{Implementation choices}
We choose to implement the algorithm in C++ because we wanted a language that offers the advantages of an OOP language, not least a faster development, that had a standard library and sufficiently ``near'' the machine to have a tighter control over the performances and the memory occupation (e.g. performances high variance due to overhead of a VM or an interpreter). We considered that absolute performances are not important here, whereas relative performances of the CRT algorithm and other MST algorithms are the focus, so finally the choice of the language is something not of the greatest importance. The IDE tool used for the implementation is JetBrains Clion.
We also used GitHub, as our code versioning control system.

Also, as already mentioned, we did use the Boost library to extend the STL that C++ already offers; we did implement FastGraph in place of BGL also to address memory issues in relation to the method Boost uses to store subgraphs of a graph; we in fact used an advanced method to store the family of subgraphs $G_i, i = 0,1,\dots,w$ that simply store the difference of vertices between them, $\Delta_{G_i} := V(G_i) - V(G_{i-1})$. It is always possibile to reconstruct every $G_i$ because this family only has \emph{induced} subgraphs, and this cost is not taken into account in the computation of the total time.\\

The main function of the implementation is in the “AlgoWEB.cpp” file.
Launching the program from this file allows us to run either the CRT algorithm or the Prim algorithm,or the Kruskal algorithm, and to view either the running time or the computed weight of the Minimum Spanning Tree. It takes some argument in input, namely the file containing the graph, a suitable value for $\varepsilon$ and the path where to save the results of the single run.

Besides, it is possible to run the random graph generator by the \texttt{grandom} utility we developed apart. It takes many arguments in input, that you can find just by calling
\begin{verbatim}
 $> grandom -h|--help
\end{verbatim}

\subsection{Random Graphs Models}
In order to give exhaustive results, we designed \texttt{grandom} to produce 4 classes of random graphs:
\begin{itemize}
\item Erd\H{o}s-R\'{e}nyi model, that builds random graphs using a uniform distribution of the edges over the vertices; his average degree $d \approx \frac{2m}{n}$;
\item Gaussian model, that builds random graphs with a ``cluster zone'' where the edges are more frequent, hence the gaussian shape of the degree distribution; we have still $d \approx \frac{2m}{n}$;
\item Barab\'{a}si-Albert model, that builds random \emph{scale-free} graphs. The average degree is $d \approx \frac{m}{n}$;
\item Watts-Strogatz model, that builds random graphs with \emph{small-world} properties and $d \approx \frac{2m}{n}$.
\end{itemize}

\label{ddd}
We want to emphasize here the fact that for the Barab\'{a}si-Albert model the average degree results to be different respect to the other models; this is due to the algorithm the scientific literature offers to build such graphs. For the other models is always possible to add an arbitrary number of edges keeping the theoretical properties valid; having, on the contrary, for the Barab\'{a}si-Albert model a certain probability $p_k$ at each step to successfully add an edge, it is not possible to build a graph with an arbitrary number of edges; the user can solely give the number of vertices. But then again, the theory states that if the algorithm starts with a complete graph of $m_0$ vertices (hence $m_0 - 1$ edges), it will produce a Barab\'{a}si-Albert graph whose average degree is scaled by this quantity. Our initial complete graph has $m_0 = 2$ vertices, so we will have $d \approx \frac{2m}{m_0 n} = \frac{m}{n}$. A little insight on the proof is the following: the distribution of the degree in the Barab\'{a}si-Albert model is a power law with a cubic exponent. Fortunately, in that case this distribution has a well defined mean. Applying a little of arithmetic we can easily see the truthfulness of what stated previously.

This difference in the models has to be bore in mind when reading the test results, because when comparing the performances over a Barab\'{a}si-Albert graph with $n$ vertices and $m$ edges and any other one of a different model with same number of vertices and edges, we will have different average degrees. Since the CRT algorithm is designed to only depend on the latter and not on $m$ nor $n$, a multiplicative factor of $2$ is to be taken into account.

Our implementation provide a \emph{memory-aware}\footnote{it calculates the average using the \emph{progressive mean} technique, avoiding arithmetic overflow} subroutine that calculates the exact value of $d$ at each run; the values obtained so far agree with the statements above.

\section{Tests}
The following section reports the results we had launching the algorithm over a variegate dataset of graphs, as described in the next paragraph.

\subsection{Dataset}
For each random graph model listed in the previous section we decided to produce a dataset; each dataset has a ``family'' of graphs that differs one from each other following a pattern for the parameters. Precisely, we composed the dataset grouping sets of graphs based on the value of the parameters, following the rules:
\begin{itemize}
\item build one set of graphs for each model (\textsf{Uniform, Gaussian, Small-World, Scale-Free});
\item every set contains in turn other sets, one for each selected value of $n$, i.e. the number of nodes ($5000$, $30000$, $55000$, $80000$, $105000$, $130000$, $155000$, $180000$, $205000$, $230000$);
\item then again, for each selected value of $d$, i.e. the average degree ($20, 100, 200$); this way we also determine the desired value for $m$,
since we have, for our models and for the fact that $m \propto n$, a proportion between the two. It's easy to see that $d \, \xrightarrow{n \rightharpoonup + \infty} \, \frac{2m}{n}$, so we have the selected values for $m$ ($10, 50, 100$ times $n$)\footnote{operatively we fixed the desired number of edges $m$, so $d$ is a random variable with mean $\frac{2m}{n}$};
\item finally, a set for each selected value of $w$, i.e. the weight ($20, 40, 60, 80$).
\end{itemize}

In conclusion we have \textsf{the number of models} $\times$ \textsf{the number of selected values for $n$} $\times$ \textsf{the number of selected values for $d$ (hence $m$)} $\times$ \textsf{the number of selected values for $w$} for a total of $4 \cdot 10 \cdot 3 \cdot 4 = 480$ graphs. This dataset, made of plain text files as already described, is quite heavy: $42.1$ GiB of data. 

\subsection{Runs}

Using the dataset just described, we used a pattern for the runs, in order to have a complete view of the behavior of the algorithm in the domain of the various parameters; every single result consists of three plots, and we obtained, inspired by the structure of the dataset:
\begin{itemize}
 \item a set of results for each model;
 \item this set containing in turn a set for each selected value of $\varepsilon$ ($0.2, 0.3, 0.4, 0.49999$);
 \item then a set for each selected value of $d/2 \simeq \frac{m}{n}$ ($10, 50, 100$);
 \item finally a set for each selected value of $w$.
\end{itemize}

The first two plots report the absolute and relative error of the CRT result compared to the correct one, calculated with Prim's algorithm; the third report the trend of the used time. As we mentioned, we indeed used Kruskal too but it's not reported here for it was not faster respect to Prim's times and unuseful for the computation of the MST weight, since already done with Prim's method.

We had this way a total of $3 \cdot 4 \cdot 3 \cdot 4 = 144$ plots, or better $48$ different cases of study. A comparison between those cases we consider meaningful finally concludes our work, as reported  below in section~\ref{sec:results}.

As mentioned before, the parameters should be selected carefully in order to provide a good approximation very fast. In this sense $\varepsilon$ plays a crucial role here: since the CRT algorithm is a probabilistic algorithm and $\varepsilon$ is indeed the driver parameter either for performances and for the accuracy of the estimate MST weight, it does not make sense to choose values too small for this parameter, as the performances could dramatically degrades (as, indeed, is expected).
So, although it could be of interest to study the right limit of $1/2$, we empirically noted that values of $\varepsilon$ below $0.2$ shows undesired behaviour of the CRT algorithm, either for the computed MST weights and for the running times.
This is not unusual dealing with theoretical algorithms that shows \emph{asymptotical} properties; the class this algorithm belongs to is known as \emph{property testing algorithms}, whose requirement is to have a query complexity much smaller than the instance size of the problem. Given that and the fact that the algorithm is not required to compute an \emph{exact} value, but to \emph{estimate} a value \emph{probabilistically near} to the correct one\footnote{iff the property we are looking for is a \emph{probabilistically checkable proof}, see~\cite{sta}}, we are not surprised that if the instance of the problem is small, the algorithm shows worse performances respect to a ``deterministic'' method. Because of this, results for $\varepsilon < 0.2$ were not comprehensive nor interesting and are not reported.

Given the intrinsically probabilistic nature of the CRT algorithm, we had to launch several runs on the same graph to have a good estimate of the CRT running time. For this purpose we decided to launch 10 runs for each graph of the dataset, and to take the average running time as the estimate of the running time; the amount of runs has been decided as a compromise between having a low variance between execution time's mode and mean values, and keeping a restrained amount of tests to do over the whole dataset. For the output value of the approximated MST weight we instead took one over the ones computed, randomly, to preserve the information on the tolerance of the estimate.

\section{Results}\label{sec:results}
 
Following there are considerations about the meaningful line charts regarding the results.

We know that the CRT time complexity is $O(dw\varepsilon^{-2} \log{\dfrac{dw}{\varepsilon}})$, where $d$ is the average degree of a graph; on the other hand, we know that the accuracy depends on $\varepsilon$ also, so we expect an inversely proportional relation with the running time. Therefore what we expect is that:
\begin{itemize}
\item by increasing one or more of the parameters $d$, $w$, there should be a worsening of the average running time;
\item keeping all the other parameters unchanged, if we consider an increase in the value of $\varepsilon$ there must be an improvement of the running time (as well as a worsening of the result, though);
\item viceversa we expect the opposite behaviour if we decrease $d$ and/or $w$ or decrease $\varepsilon$ with the other parameters unchanged.
\end{itemize}
Let us call the above \emph{crucial parameters}.

What we still need to know is what happens to the error; we should expect a direct proportion with the running times, so the above considerations could be also valid for the error. On the contrary, we'll see that this is not exactly respected.

\begin{figure}[htbp]
 \centering
 \subfloat[][\emph{Running time; for completeness, Kruskal's time is also reported}.]
 {\includegraphics[width=.8\textwidth]{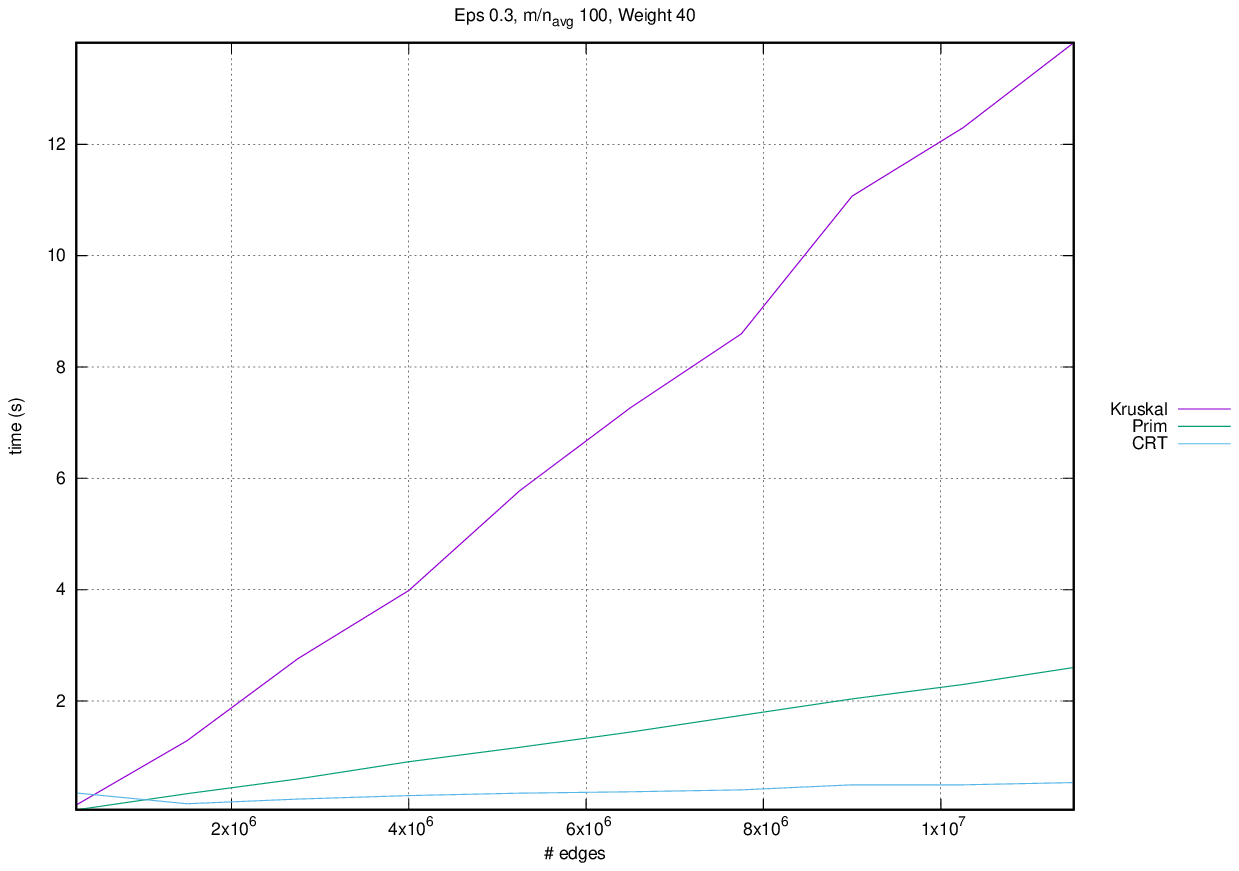}\label{U_03_50_40_time_kruskal}} \\
 \subfloat[][\emph{Relative error}.]
 {\includegraphics[width=.45\textwidth]{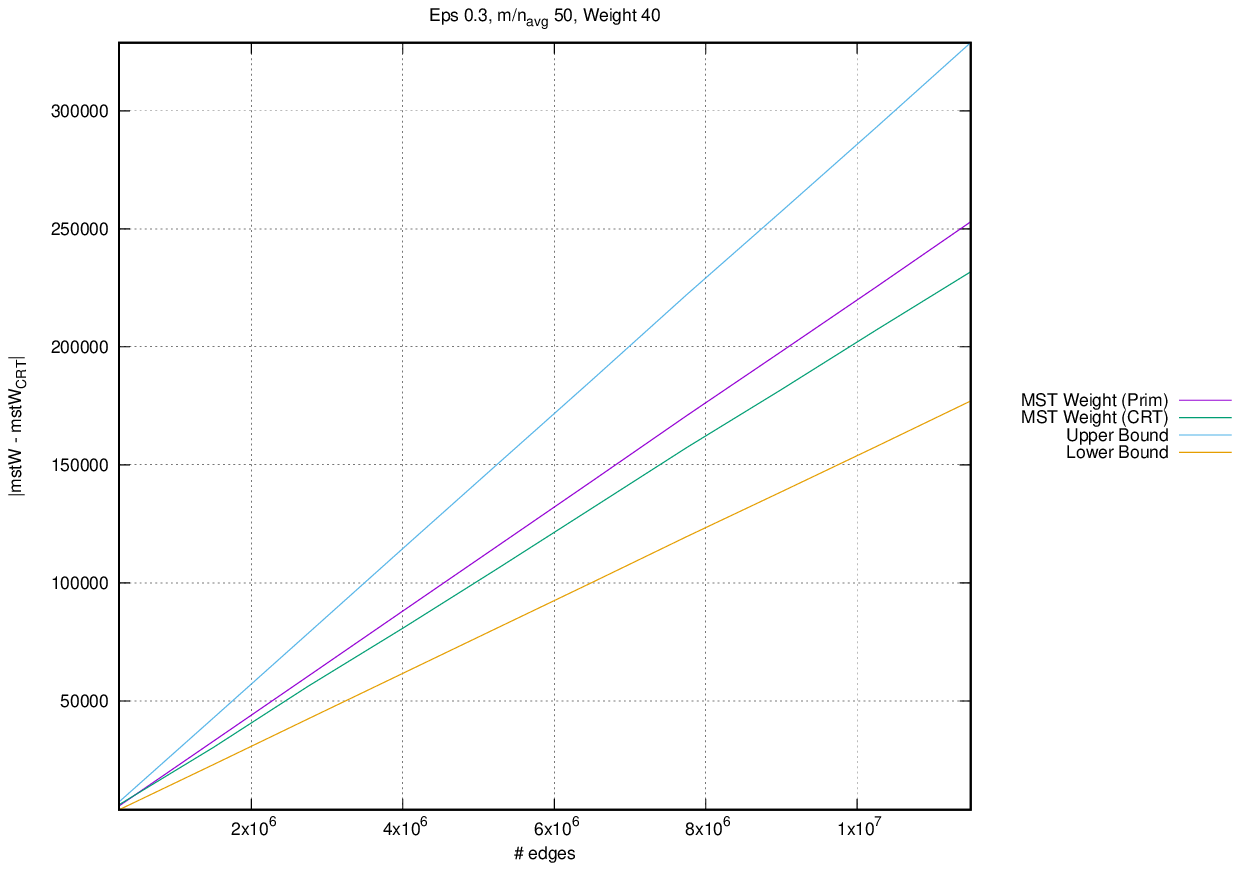}\label{U_03_50_40_abs}} \quad
 \subfloat[][\emph{Absolute error}.]
 {\includegraphics[width=.45\textwidth]{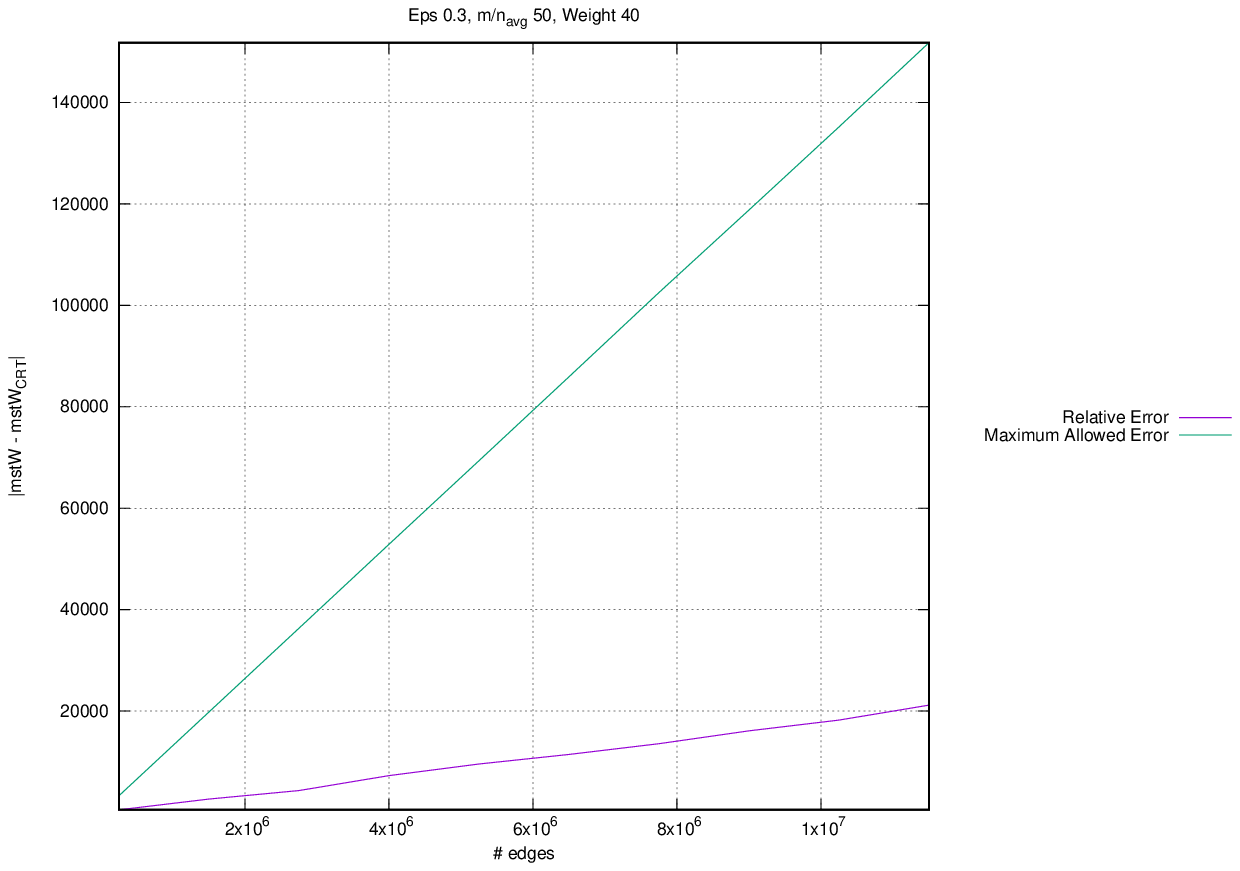}\label{U_03_50_40_rel}}
 \caption{Sublinearity of the CRT algorithm and related error trends.}
 \label{is_sublin}
\end{figure}

First thing to show is that the CRT algorithm is sublinear in the number of edges, hence is better than any other \emph{exact} algorithm. This can be easily seen in figures from~\ref{U_03_50_40_time_kruskal} to~\ref{U_03_50_40_rel}. For the rest of the plots listed from now on it is possibile to see, above each graph, the values of the parameters for the presented run.

It is interesting to note that the correct value, computed with Prim's algorithm, it's linear in the number of edges (figure~\ref{U_03_50_40_abs}); we want to stress the fact that this is not the general case, and that this trend is due to the uniform distribution of the weights over the edges. We will dissect this point more in section~\ref{rtn}.

Given that the CRT Algorithm respects the sub-linearity constraint, let's now see the variations that occur when selectively changing other parameters. For the sake of completeness, in figures~\ref{is_sublin} informations about Kruskal's runs are reported, yet we won't report them in the charts that follow.

\subsection{Variations of crucial parameters}

\subsubsection{Average degree $d$}
Let us now see the behaviour for the variation of $d$; we will initially concentrate on the running time and solely for the \textsf{uniform} model. The selected values of $\varepsilon$ and $w$ are respectivey $0.3$ and $40$. 

\begin{figure}[htbp]
 \centering
 \subfloat[][$d \simeq 20$]
 {\includegraphics[width=.65\textwidth]{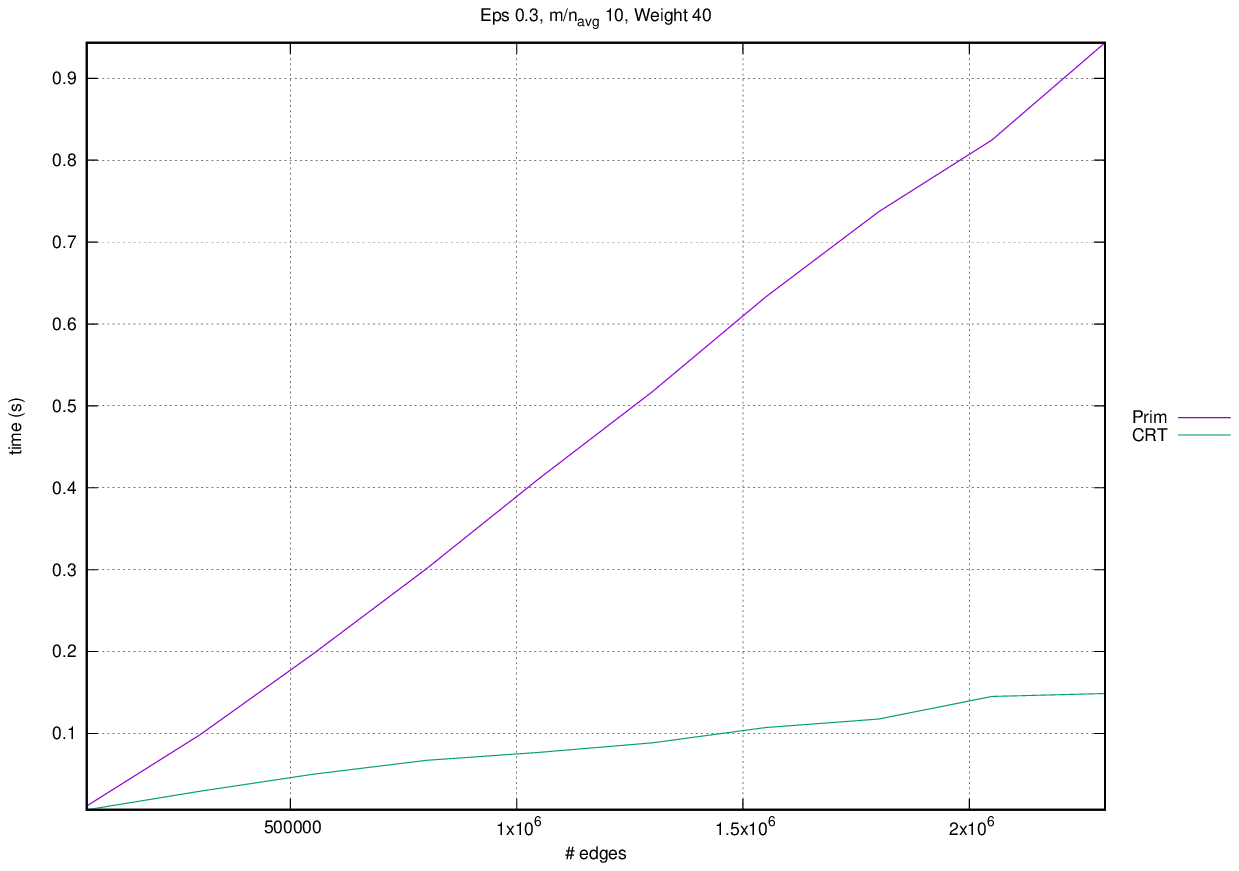}\label{U_03_10_40_time}} \\
 \subfloat[][$d \simeq 100$]
 {\includegraphics[width=.65\textwidth]{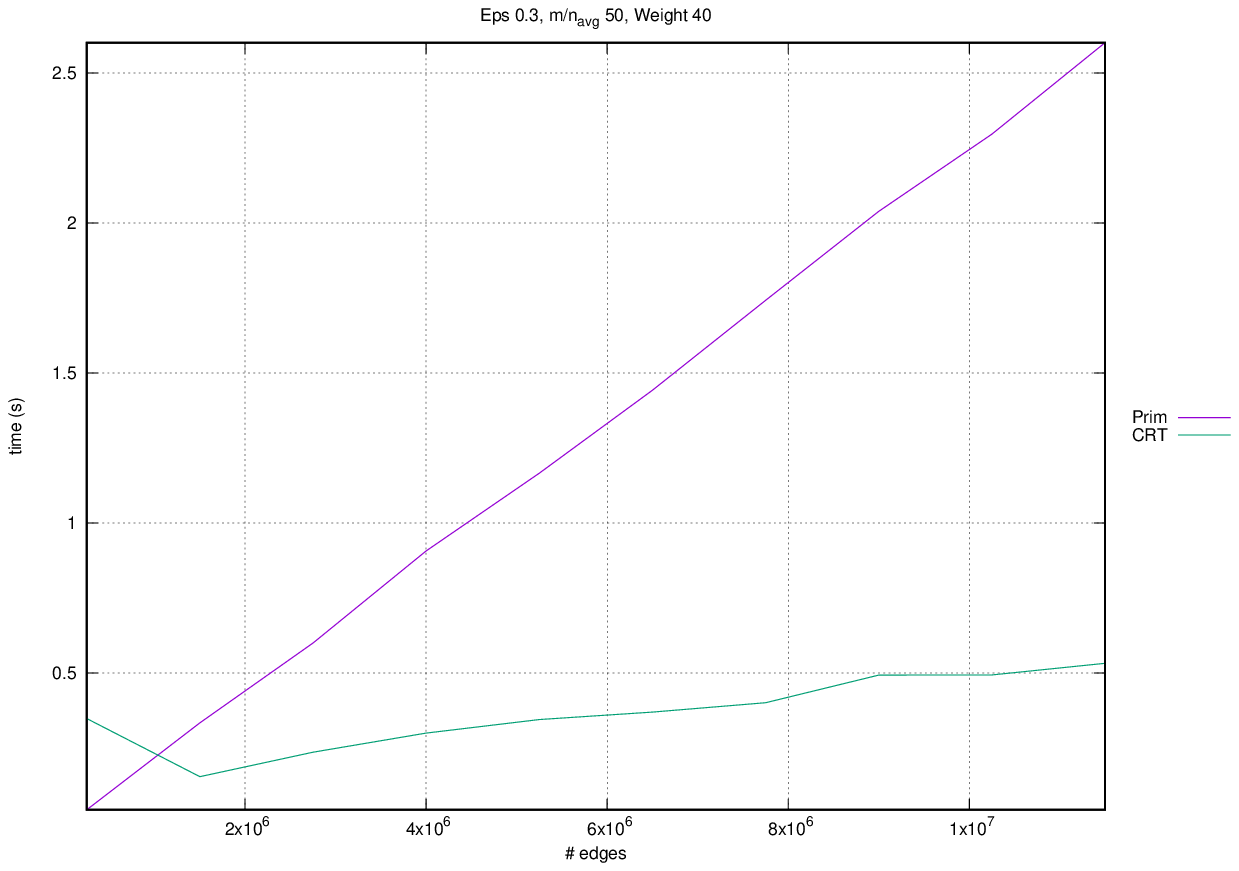}\label{U_03_50_40_time}} \\
 \subfloat[][$d \simeq 200$]
 {\includegraphics[width=.65\textwidth]{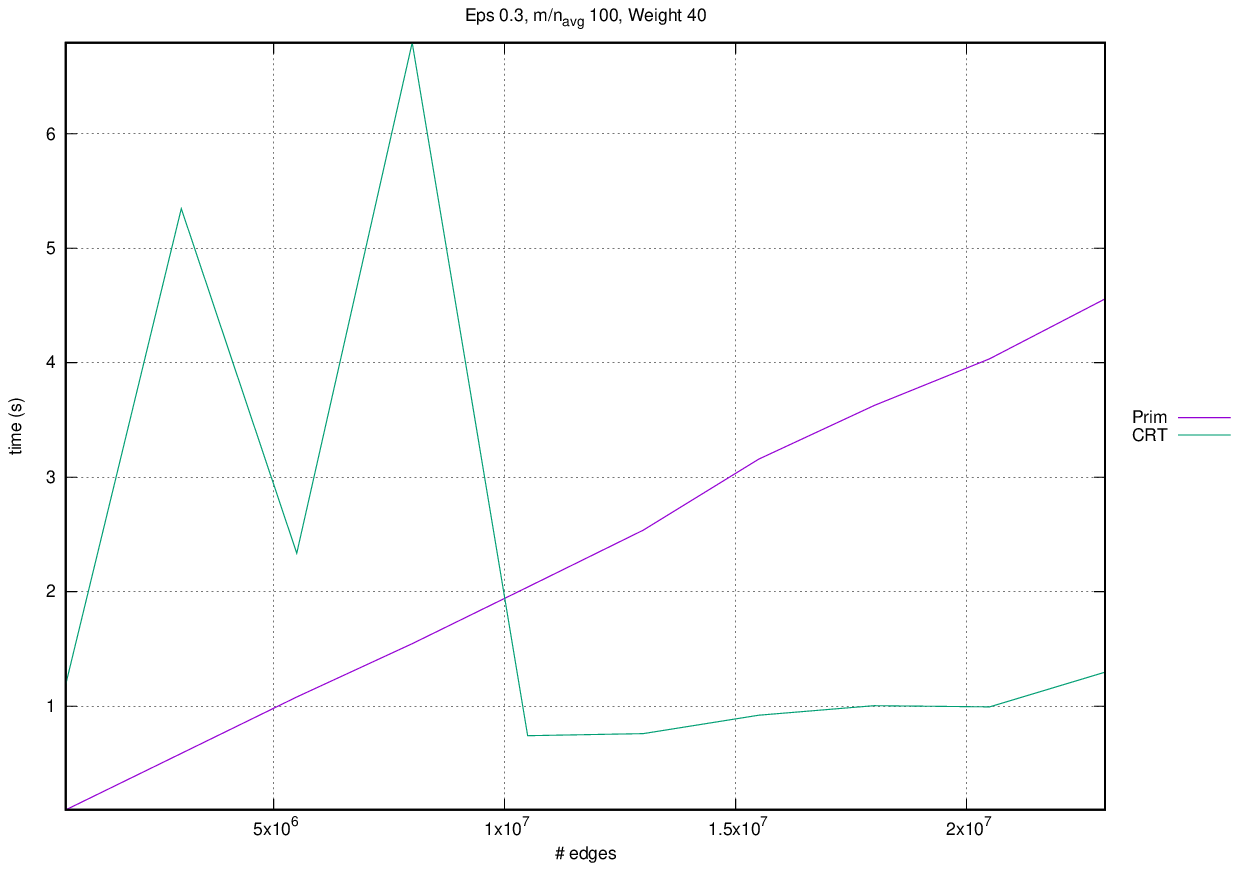}\label{U_03_100_40_time}}
 \caption{Behaviour for the increase of $d$. ({\sf Uniform model})}
 \label{d_increase_time}
\end{figure}

As we can see in figures~\ref{U_03_10_40_time} to~\ref{U_03_100_40_time}, there is a worsening in the performance \emph{for small instances}: an increase of the average degree $d$ is somewhat correlated to a loss of performances to the point that our property testing algorithm needs more time that the deterministic one; still, that seems to be true \emph{under} a certain dimension of the instance of the graph, so that we don't lose the truthfulness of the theoretical time complexity because for a fixed $d^*$ it will always be possibile to find empirically a certain number of edges $m^* \propto d^*$ beyond which the running time function is always below $C \cdot dw\varepsilon^{-2} \log{\dfrac{dw}{\varepsilon}}$ for a certain $C$.\footnote{that comes directly from the definition of \emph{asymptotic complexity}}
We want here to highlight the crucial fact that the algorithm behaves better on \emph{big} instances, where with ``\emph{big}'' we refer to the parameters the performances depend on, namely $d$, $w$. Almost all the trends reported in this paper, in fact, show this initial ``bad'' curve and, from a certain value onward, a sublinear behaviour. We will discuss further about this in section~\vref{gbb}.

\begin{figure}[htbp]
 \centering
 \subfloat[][$d \simeq 20$]
 {\includegraphics[width=.65\textwidth]{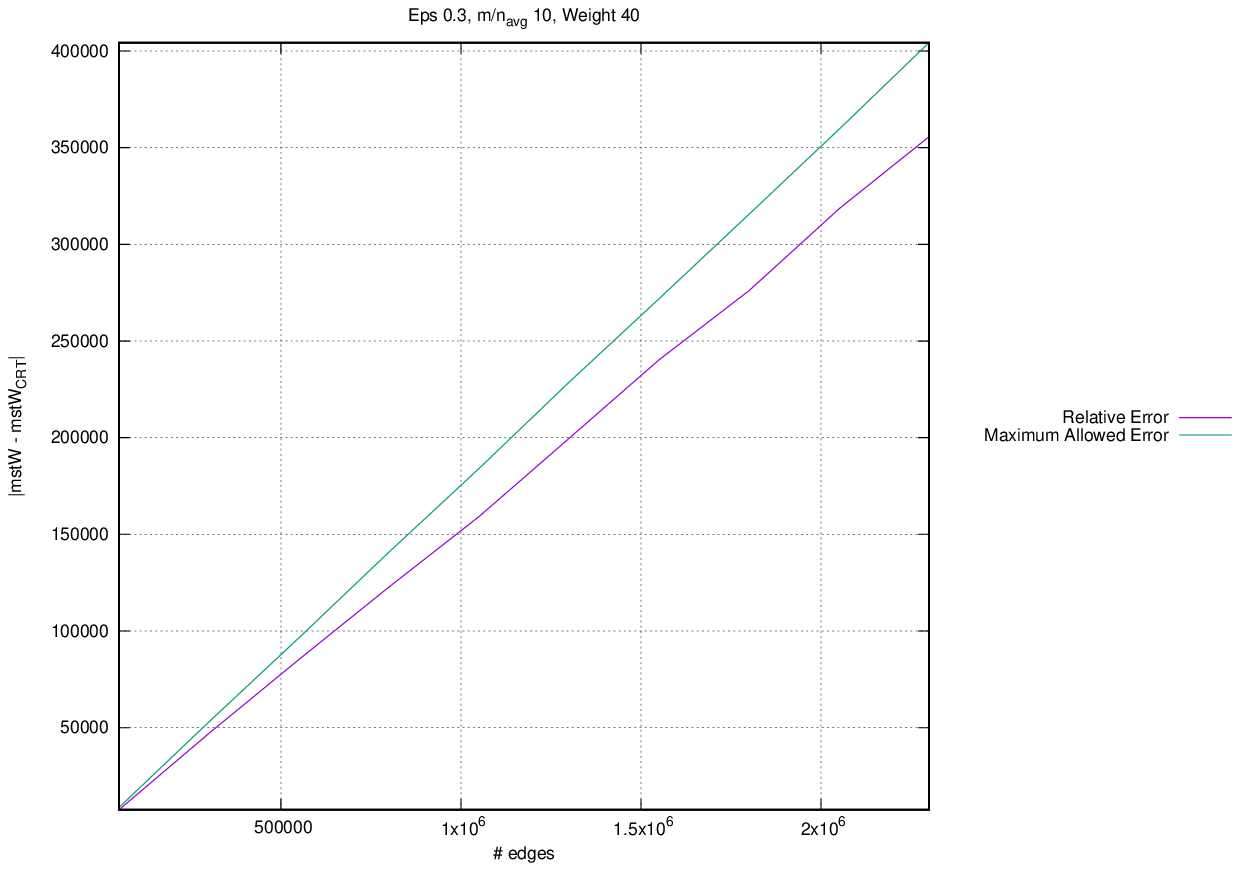}\label{U_03_10_40_rel}} \\
 \subfloat[][$d \simeq 100$]
 {\includegraphics[width=.65\textwidth]{plots/uniform_03_50_40_rel}\label{U_03_50_40_rel2}} \\
 \subfloat[][$d \simeq 200$]
 {\includegraphics[width=.65\textwidth]{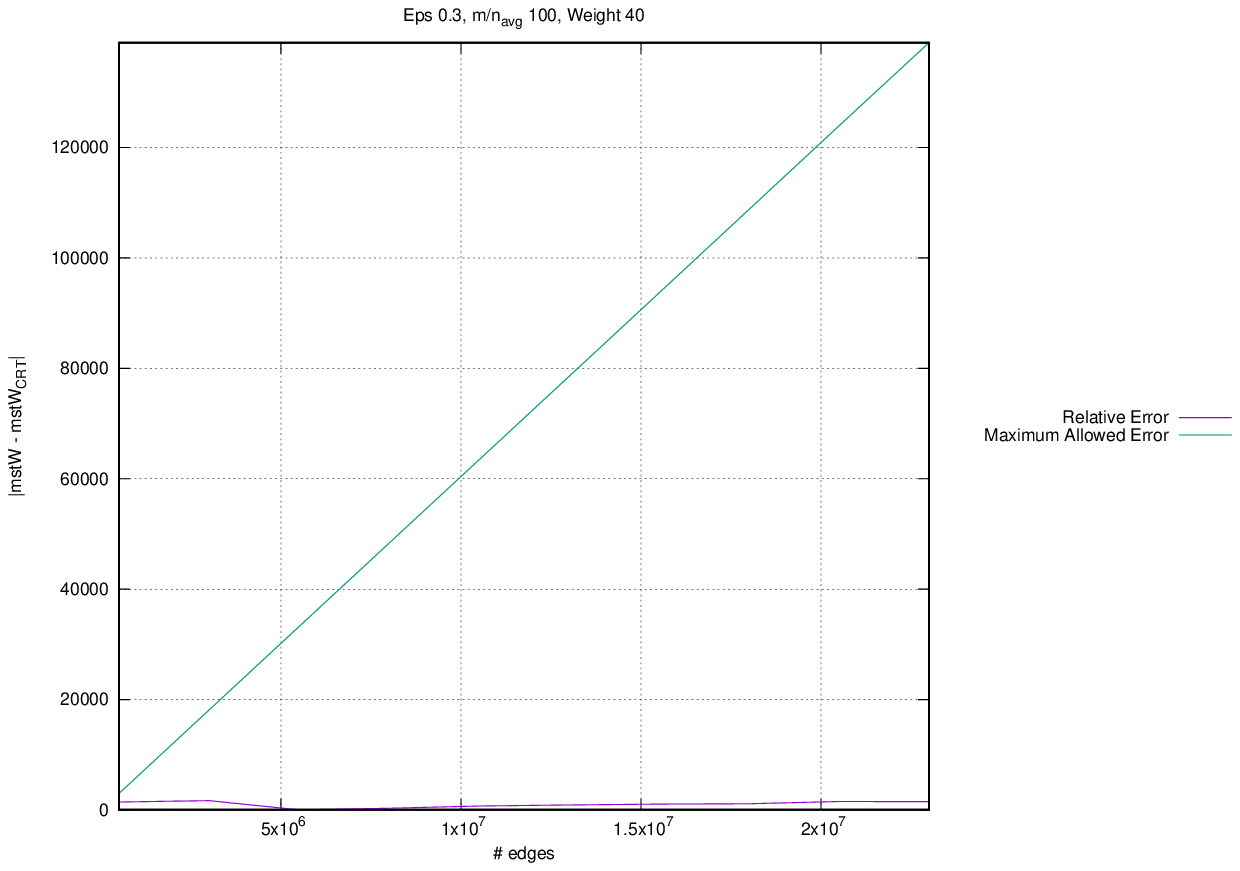}\label{U_03_100_40_rel}}
 \caption{Error behaviour for the increase of $d$. ({\sf Uniform model})}
 \label{d_increase_rel}
\end{figure}

We would speculate that the error could be related to this. Indeed in figure~\ref{d_increase_rel} we can see that to an increase of $d$ corresponds a dramatic annihilation of the error; to explain this point we use a simplified example. Let us consider a minimal complete graph\footnote{a connected graph with the least number of edges, i.e. $n-1$}; the algorithm launches some BFSes on a \emph{strict subset} of the graph, in more steps that could be interrupted according to certain rules, based in part on a stochastic process. It is easily provable that in this kind of graphs we have a worst case of $n-1$ hops between to vertices $u$ and $v$; if we add a random edge to this graph, the length of this path decrease \emph{at least} of one hop. By induction we can hence prove that the \emph{diameter} of the graph decreases as $\left| E(G) \right|$ grows. In other words having a stronger connection within the graph (i.e. a greater probability to visit a vertex $v$ from $u$ in $k$ hops of a random walk) increases the probability to have a complete view of the graph, that is more \emph{information} about the MST weight.

Moreover, we saw in our study of all the results showed in this paper, that the performances of the CRT are completely untied from the number of vertices $n$ and from the number of edges $m$ of the input graph; this suggests us also that the error is in turn driven solely by the parameters responsible of the algorithm's complexity, as the results that follow are in fact going to prove.

\begin{figure}[htbp]
 \centering
 \subfloat[][gaussian, $d \simeq 20$]
 {\includegraphics[width=.65\textwidth]{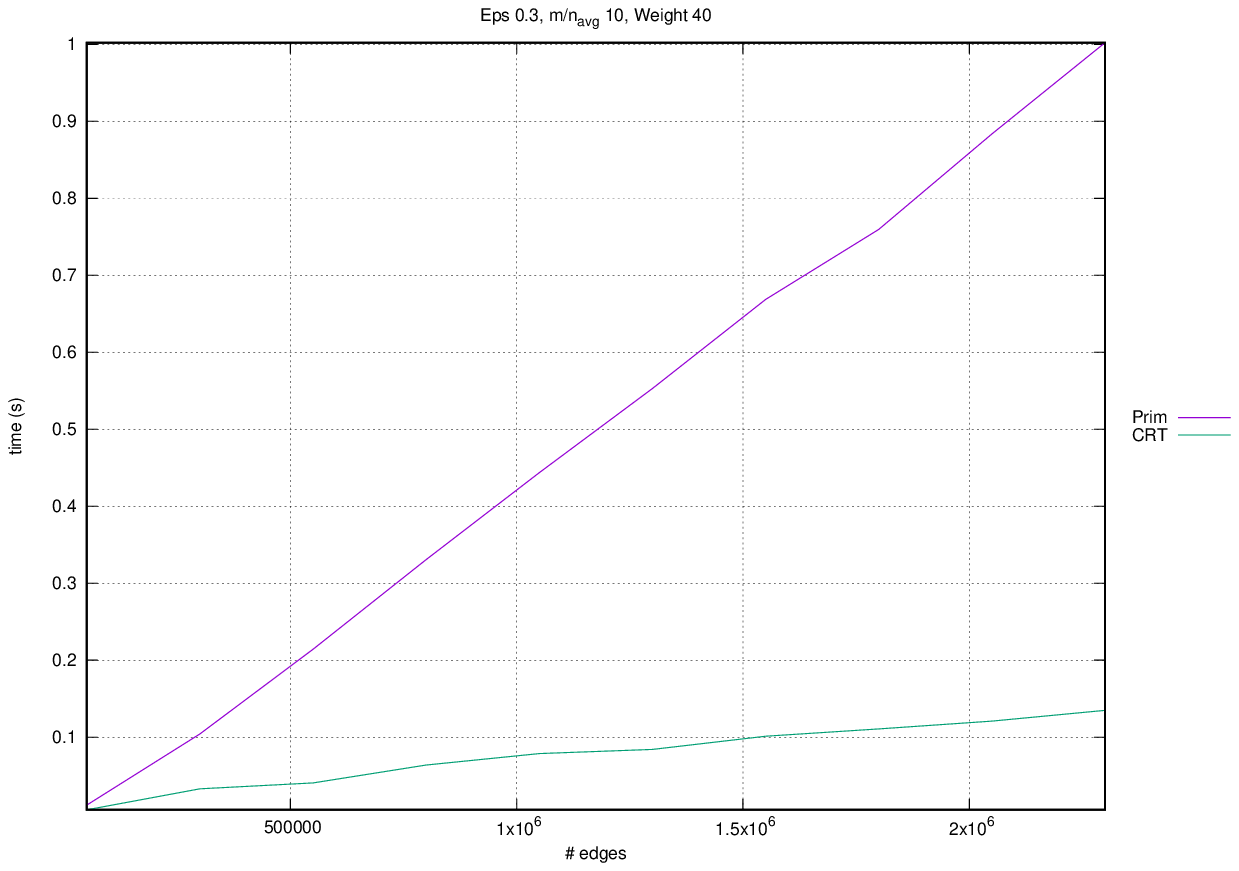}\label{G_03_10_40_time}}
 \subfloat[][small-world, $d \simeq 20$]
 {\includegraphics[width=.65\textwidth]{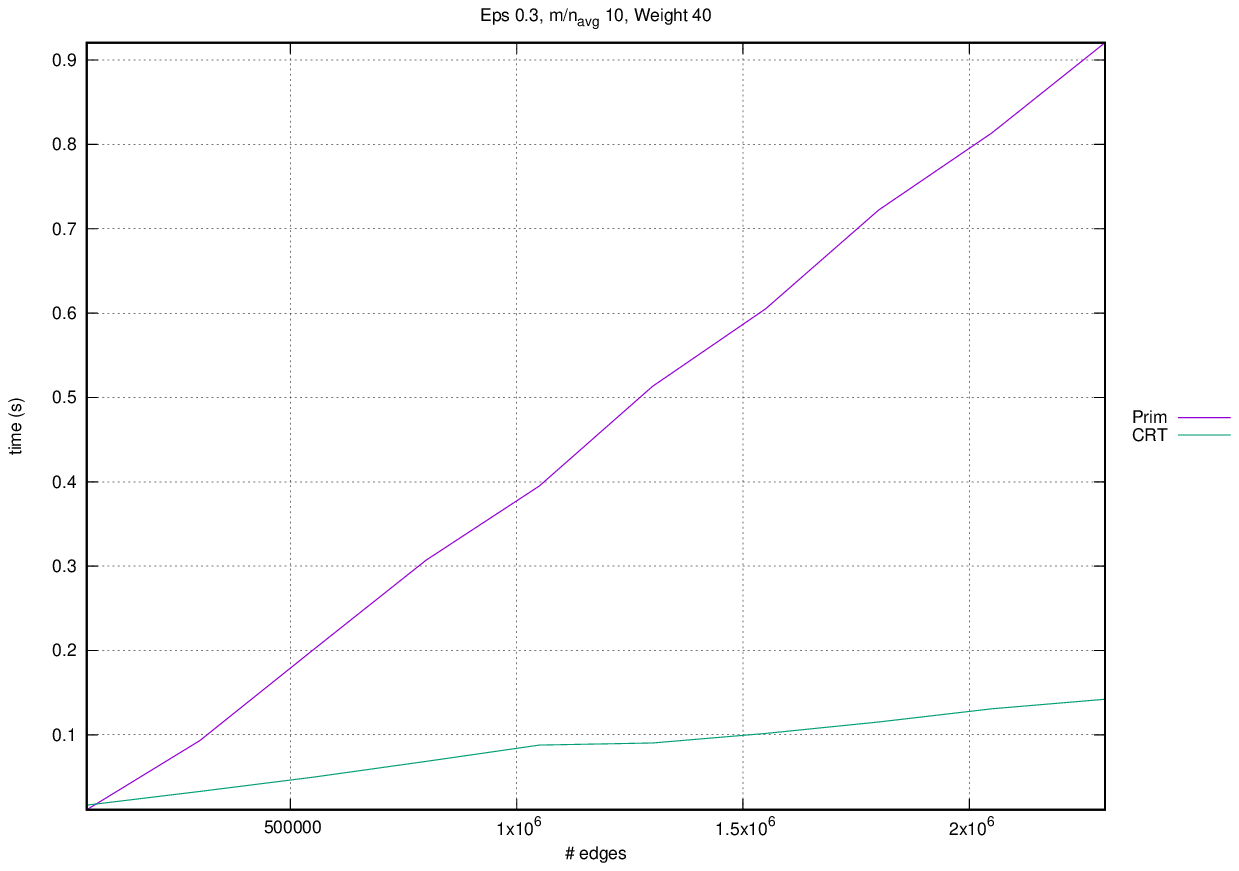}\label{SM_03_10_40_time}} \\
 \subfloat[][gaussian, $d \simeq 100$]
 {\includegraphics[width=.65\textwidth]{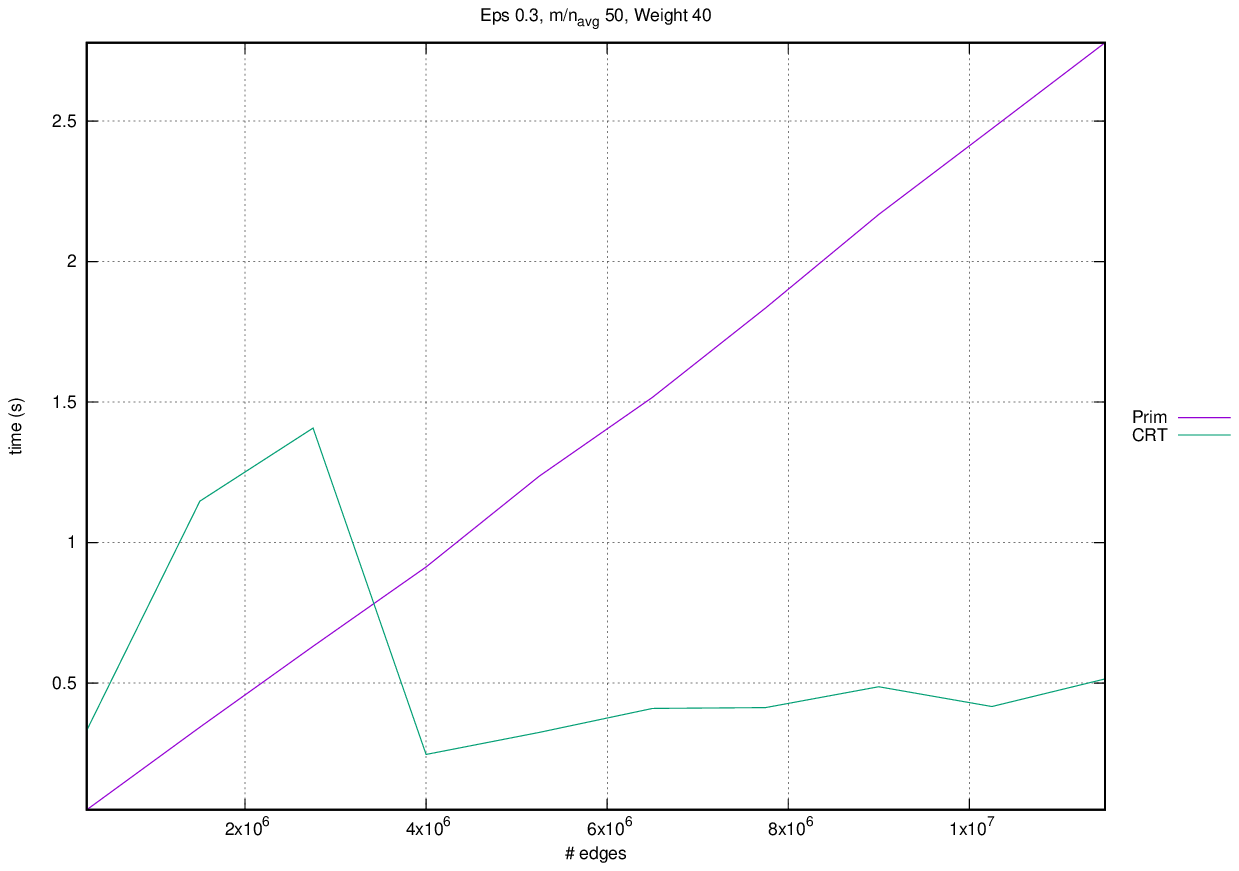}\label{G_03_50_40_time}}
 \subfloat[][small-world, $d \simeq 100$]
 {\includegraphics[width=.65\textwidth]{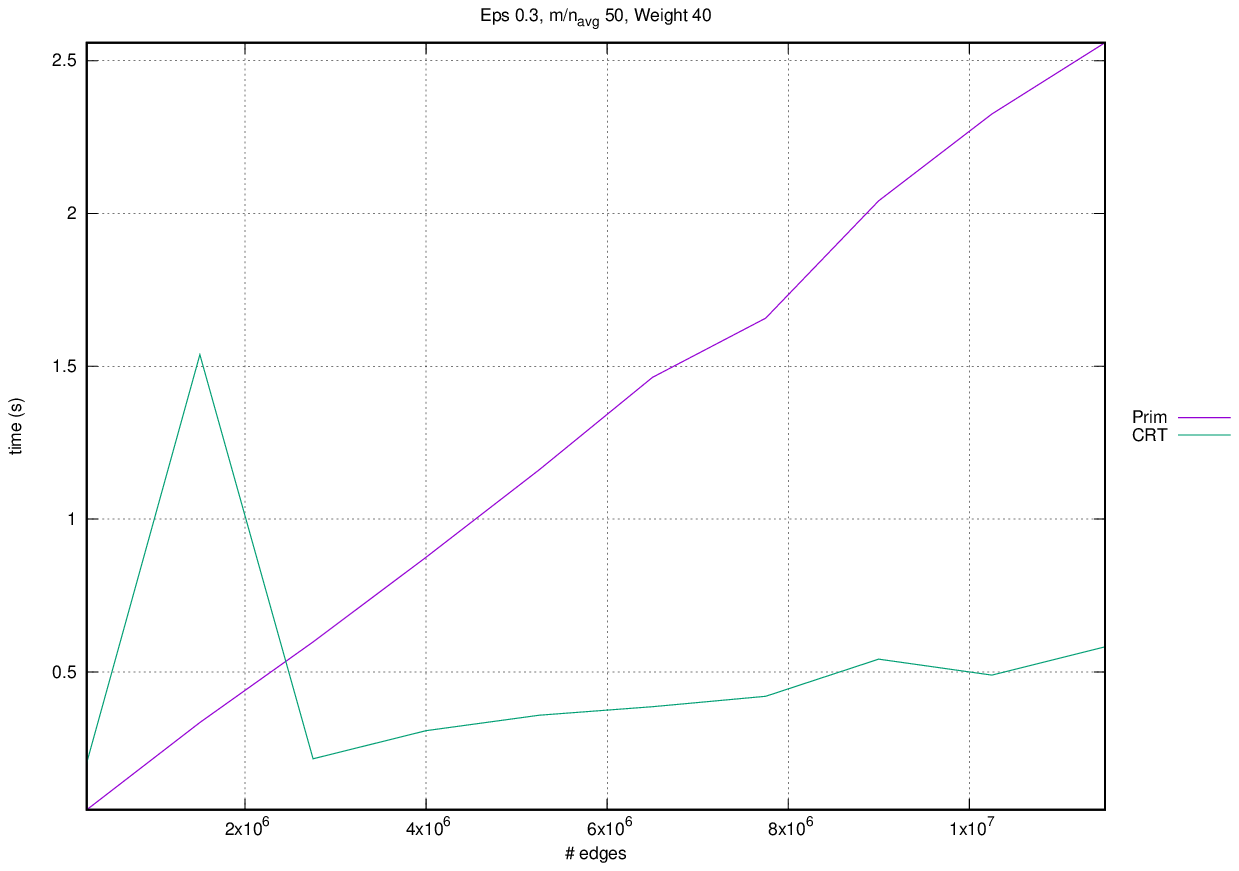}\label{SM_03_50_40_time}} \\
 \subfloat[][gaussian, $d \simeq 200$]
 {\includegraphics[width=.65\textwidth]{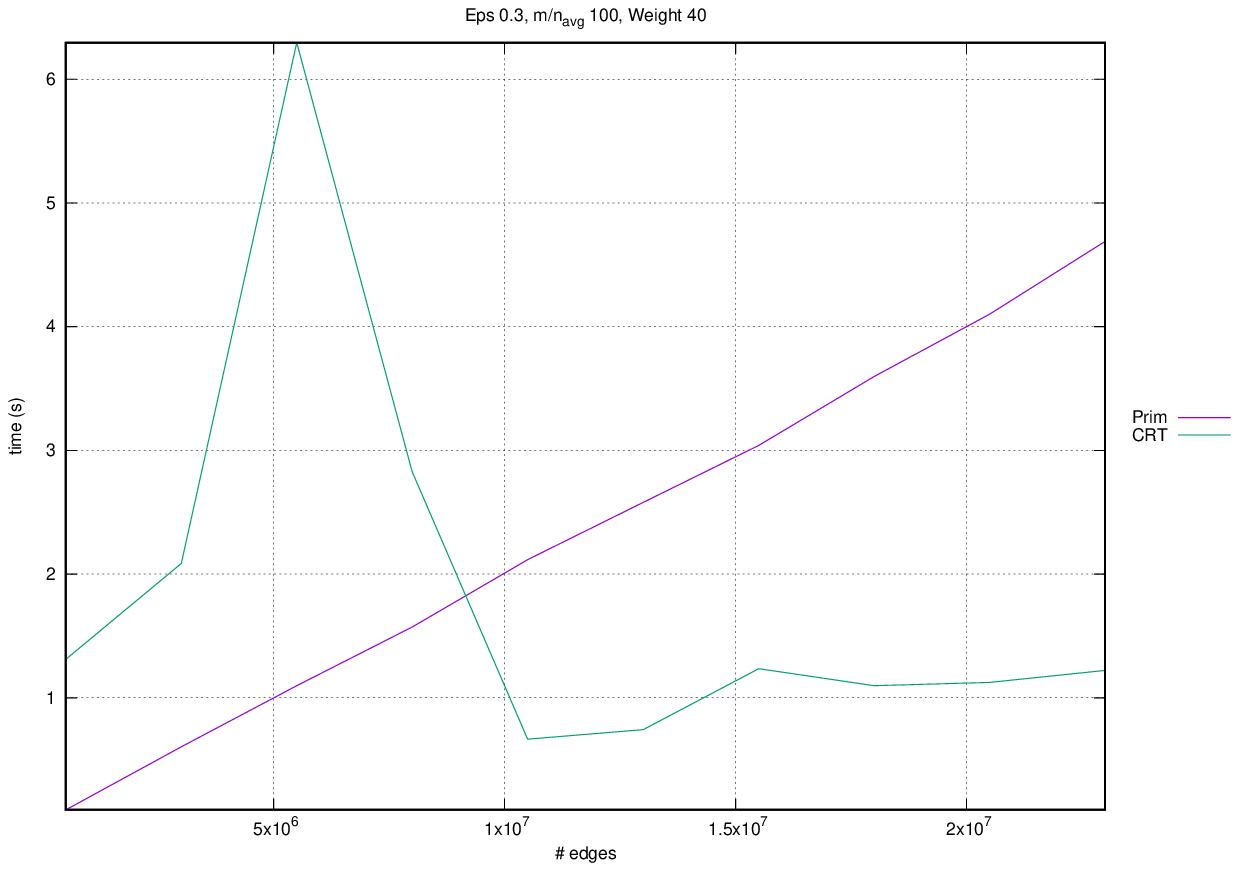}\label{G_03_100_40_time}}
 \subfloat[][small-world, $d \simeq 200$]
 {\includegraphics[width=.65\textwidth]{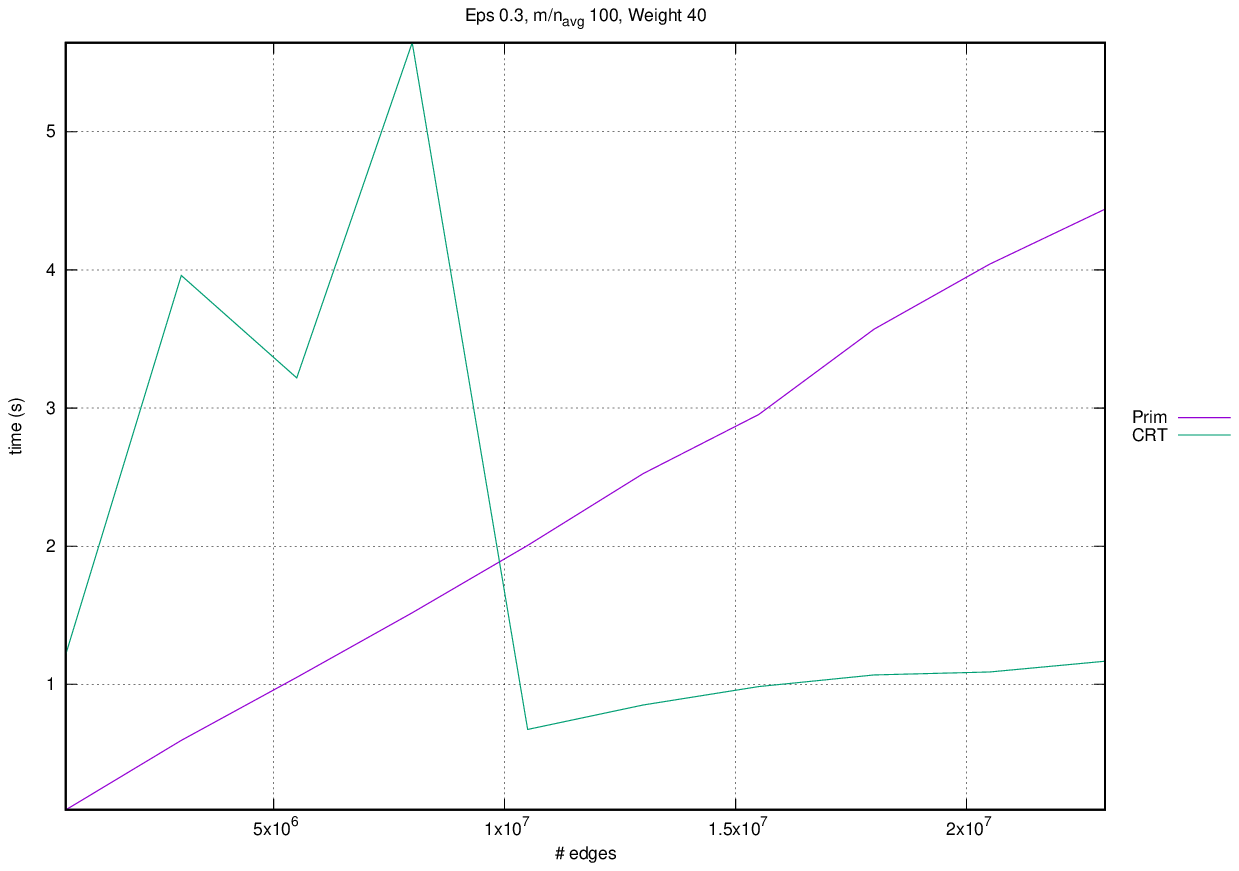}\label{SM_03_100_40_time}}
 \caption{Execution times behaviour for the increase of $d$, different models.}
 \label{d_increase_time2}
\end{figure}

In figure~\ref{d_increase_time2} we summarize the results for the Gaussian and Small-World models, noticing that they equate the one we showed about the uniform model. This suggests that the algorithm complexity does not depend on the dimension and clustering coefficient of the graphs, being those the main differences from one model to another.

\begin{figure}[htbp]
 \centering
 \subfloat[][gaussian, $d \simeq 20$]
 {\includegraphics[width=.65\textwidth]{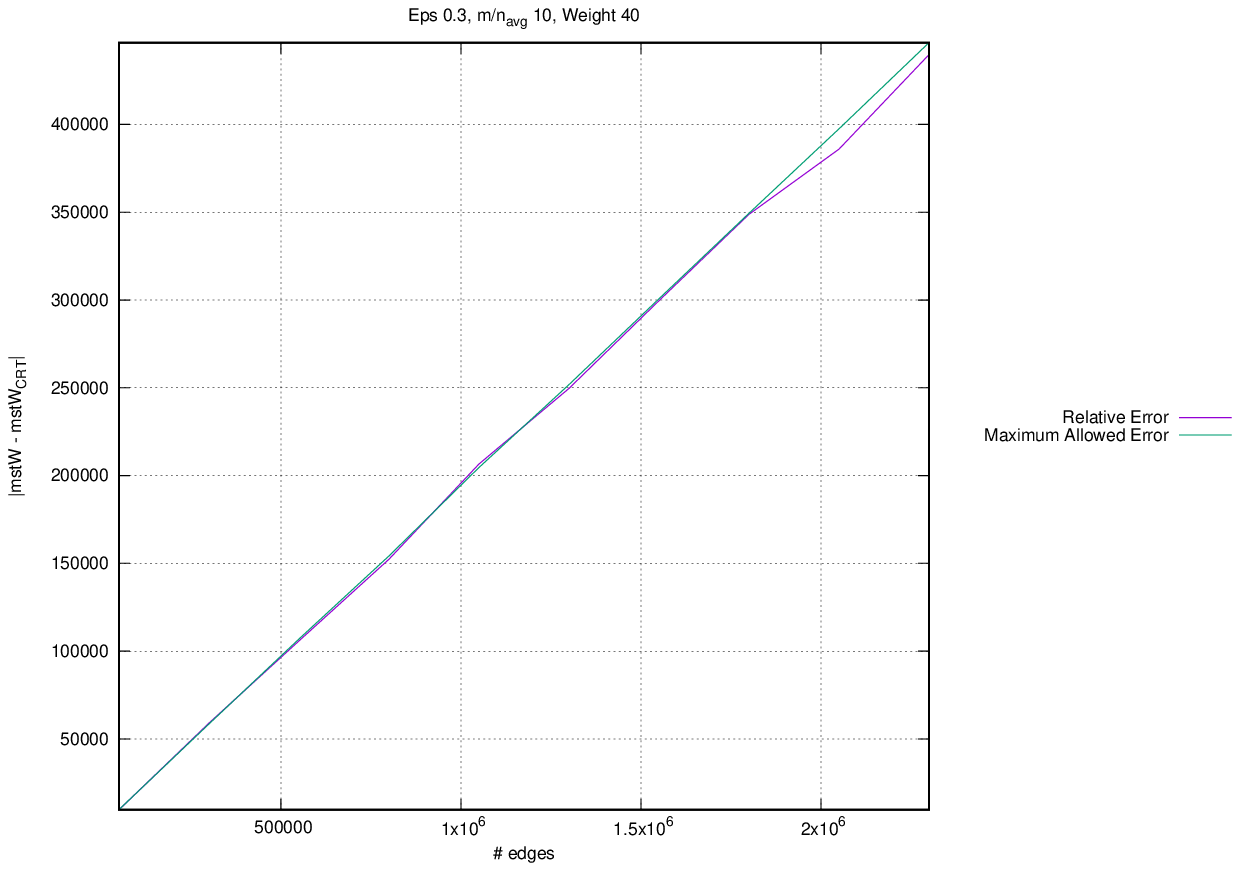}\label{G_03_10_40_rel}}
 \subfloat[][small-world, $d \simeq 20$]
 {\includegraphics[width=.65\textwidth]{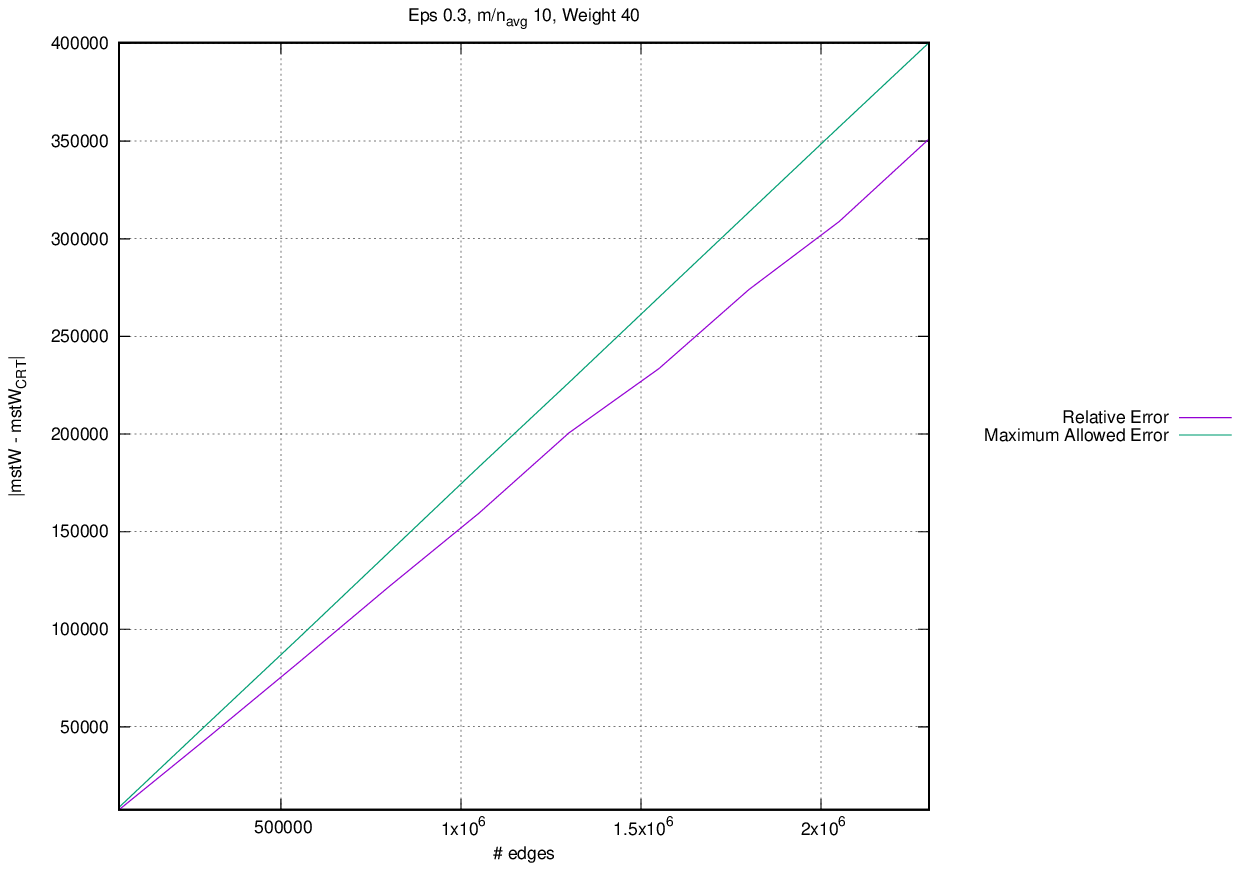}\label{SM_03_10_40_rel}} \\
 \subfloat[][gaussian, $d \simeq 100$]
 {\includegraphics[width=.65\textwidth]{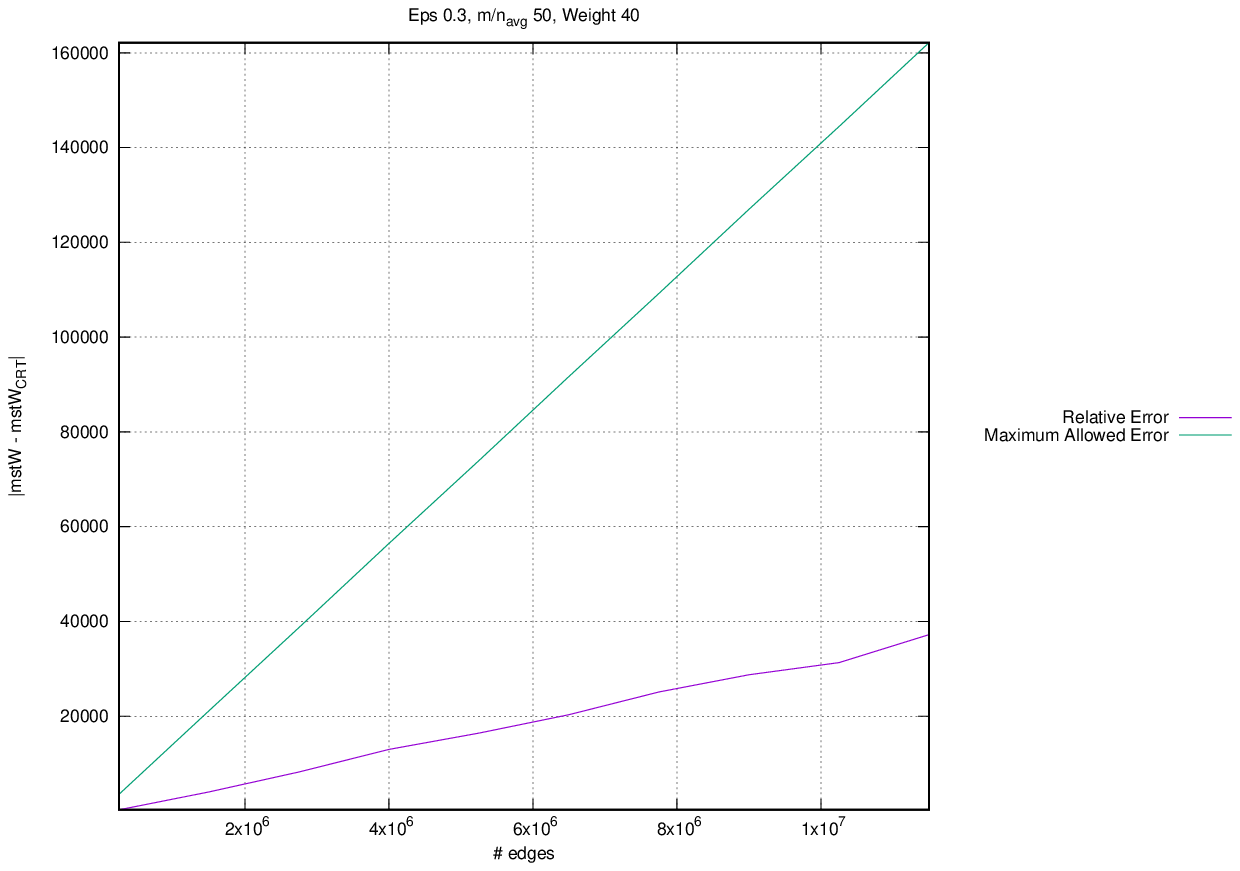}\label{G_03_50_40_rel}}
 \subfloat[][small-world, $d \simeq 100$]
 {\includegraphics[width=.65\textwidth]{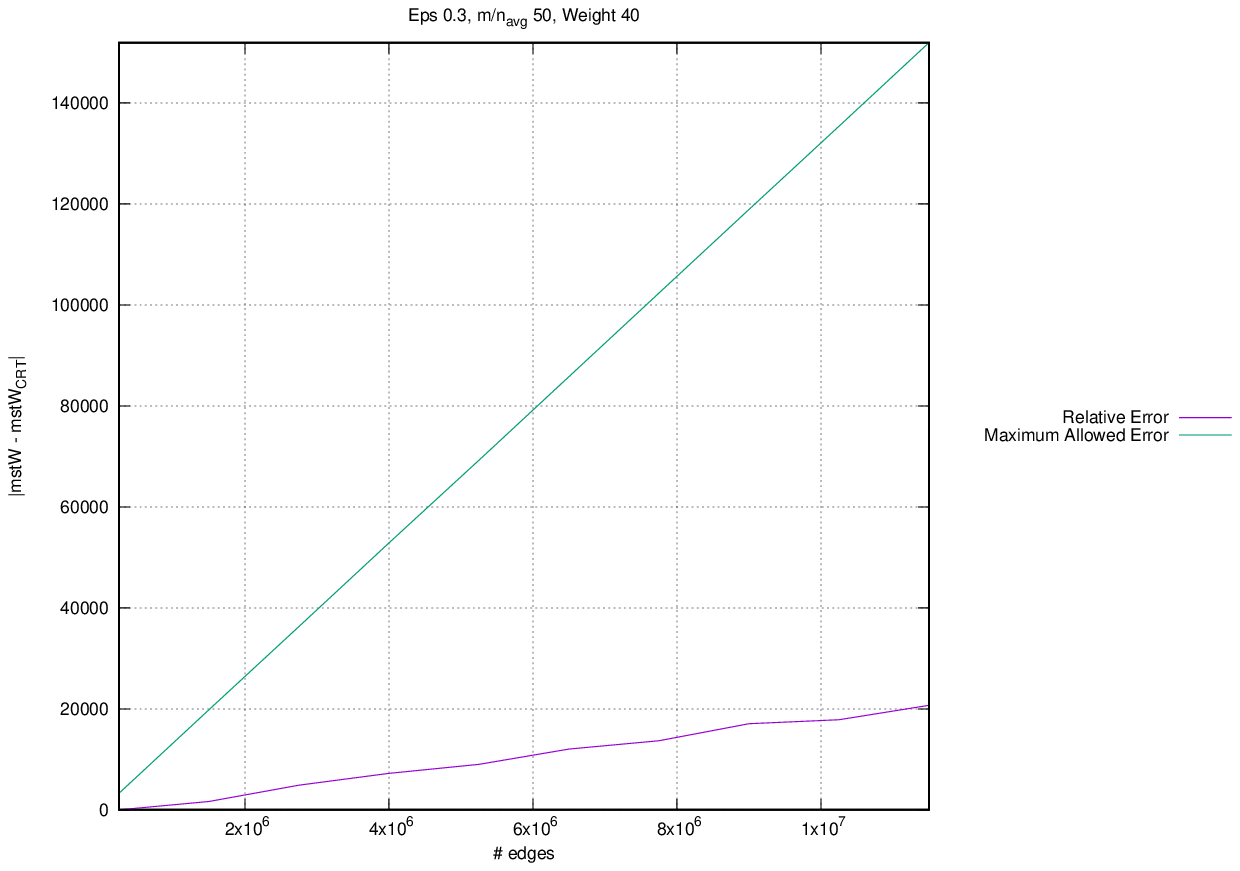}\label{SM_03_50_40_rel}} \\
 \subfloat[][gaussian, $d \simeq 200$]
 {\includegraphics[width=.65\textwidth]{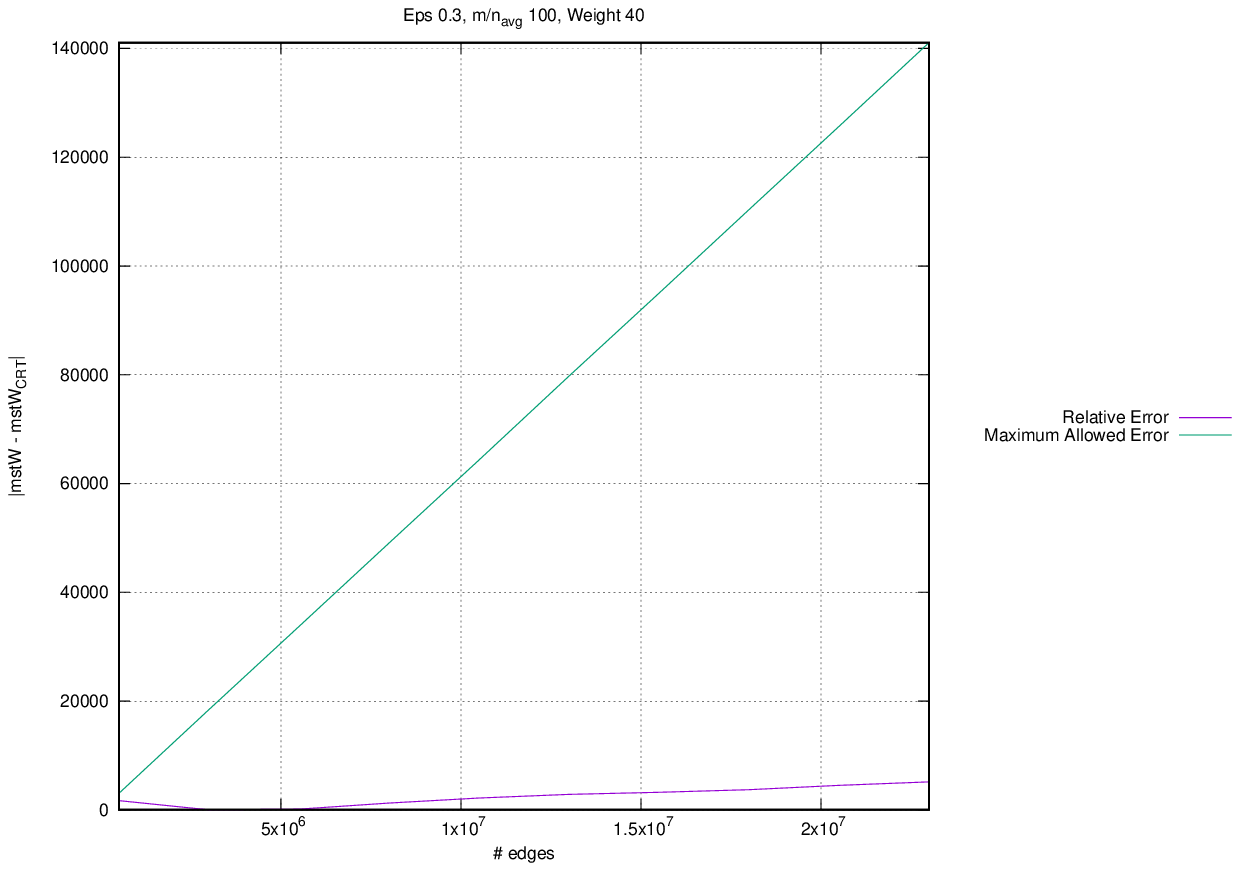}\label{G_03_100_40_rel}}
 \subfloat[][small-world, $d \simeq 200$]
 {\includegraphics[width=.65\textwidth]{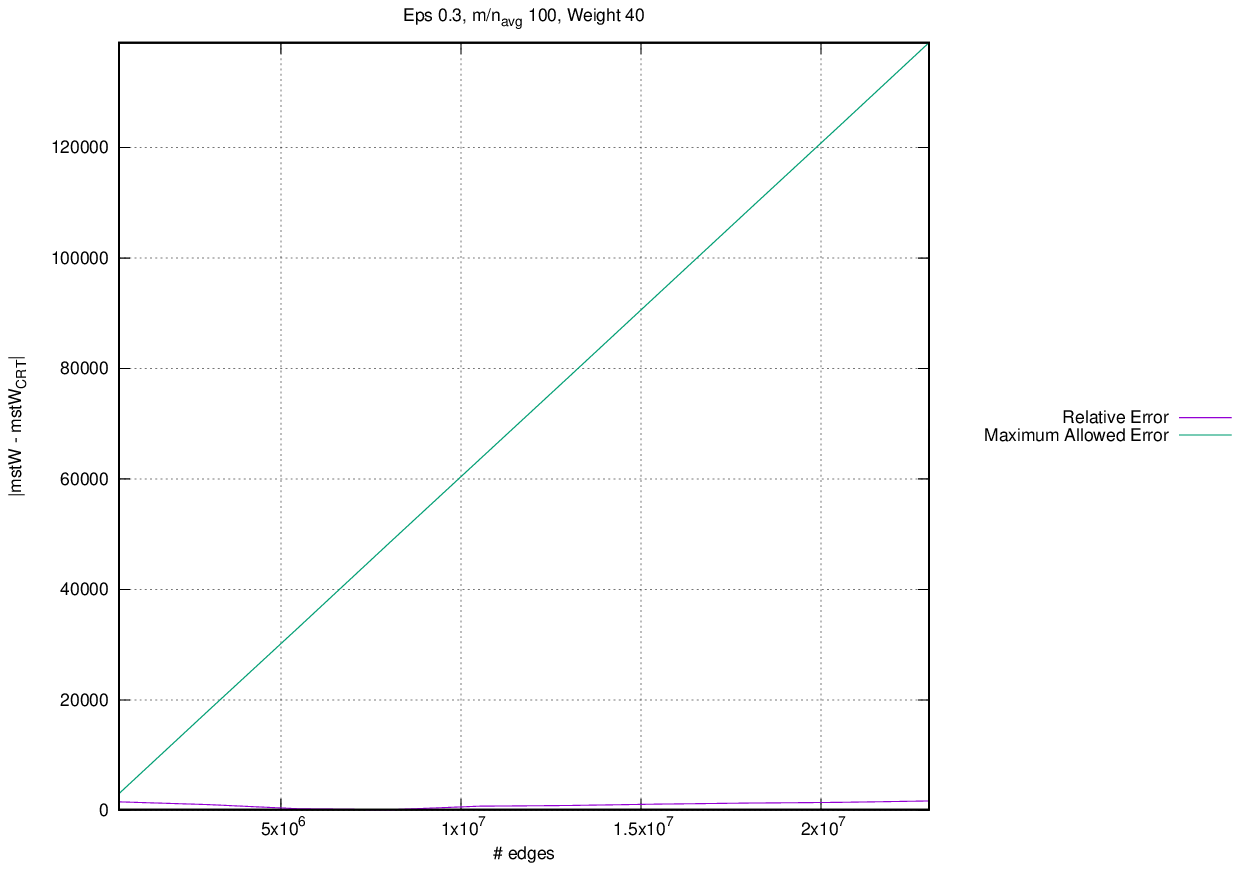}\label{SM_03_100_40_rel}}
 \caption{Error behaviour for the increase of $d$, different models.}
 \label{d_increase_rel2}
\end{figure}

In figure~\ref{d_increase_rel2} we summarize instead the trend of the relative error; we see here a slightly different evolution. We cannot conclude, as we did for the time complexity, that the error doesn't suffer from the different graph model. The error in fact depends on the clustering coefficient, because it is going to grow dependently on the number of not accomplished BFSes: each one of them cause in fact a loss of information. The algorithm, as well explained in~\cite{crt}, during the BFSes phase avoid to explore nodes that shows a high degree and even stops when encounters hubs.

In other words, having equals values of $d$ in two different runs of the algorithm, we see that its time complexity trend remain the same; assuming that $\delta$ is the sample mean of the vertices degree, this tells us that the time complexity is bound to the \emph{average} of $\delta$ and, since we have a growth of the error on graphs that contains hubs, we also conclude that the relative error is bound to the \emph{variance} of $\delta$.

\subsection{Maximum weight $w$}
\begin{figure}[htbp]
 \centering
 \subfloat[][$w = 20$]
 {\includegraphics[width=.65\textwidth]{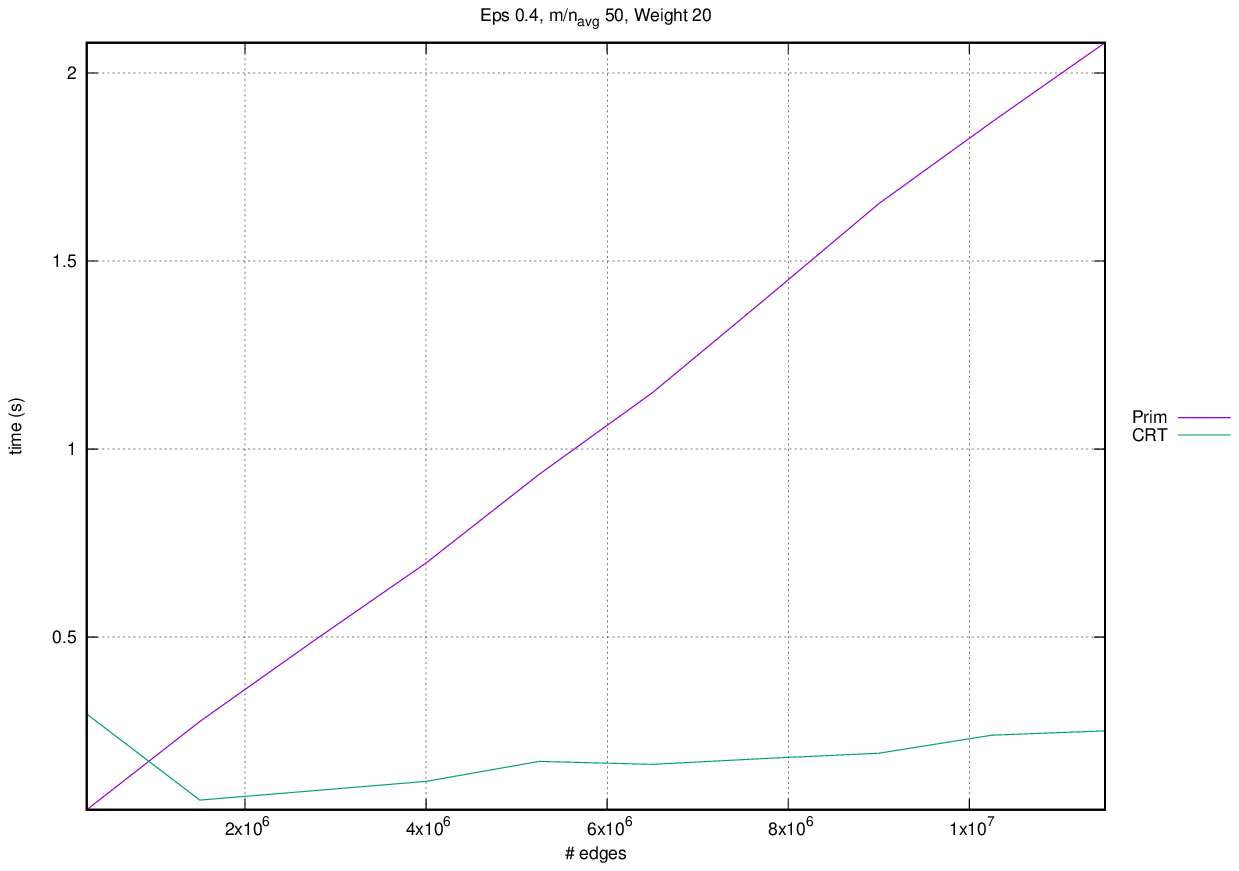}\label{U_04_50_20_time}}
 \subfloat[][$w = 40$]
 {\includegraphics[width=.65\textwidth]{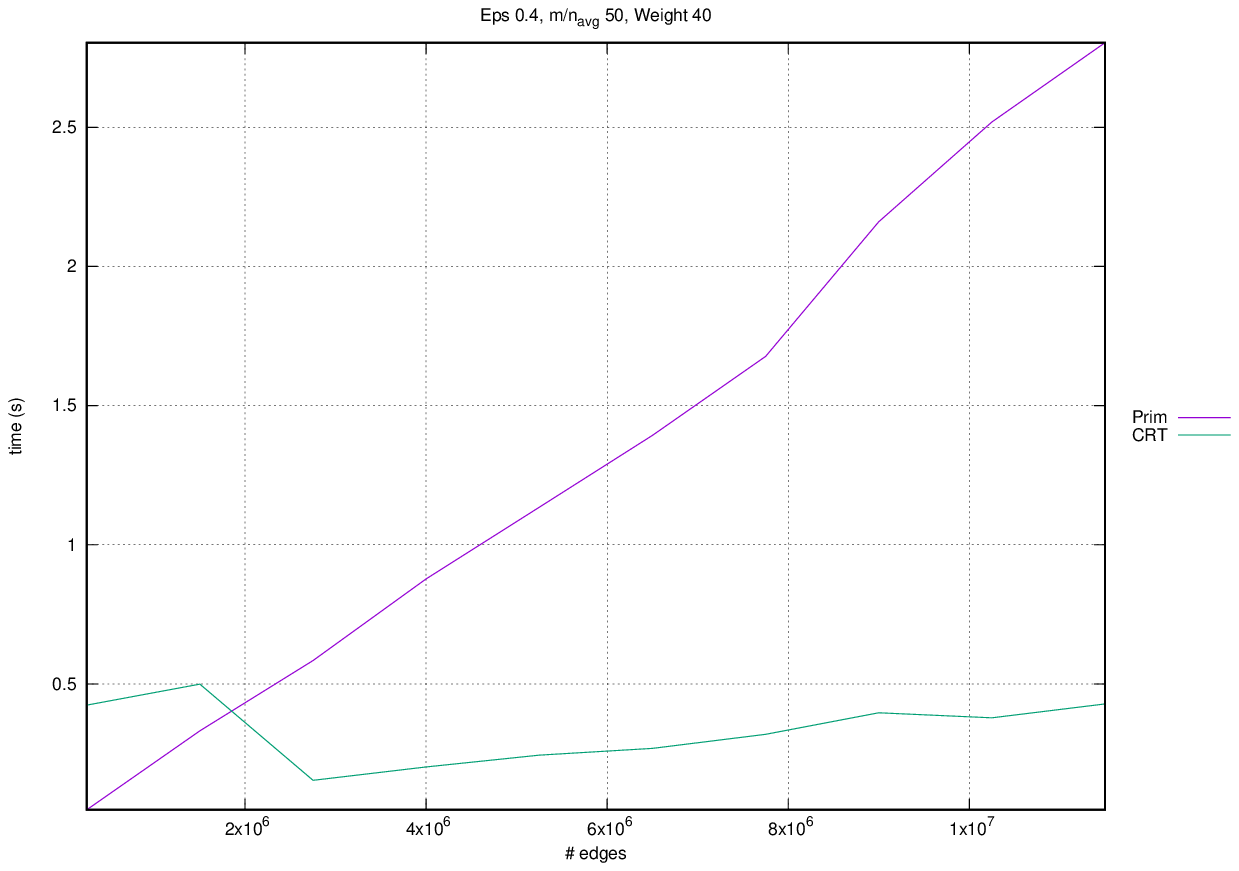}\label{U_04_50_40_time}} \\
 \subfloat[][$w = 60$]
 {\includegraphics[width=.65\textwidth]{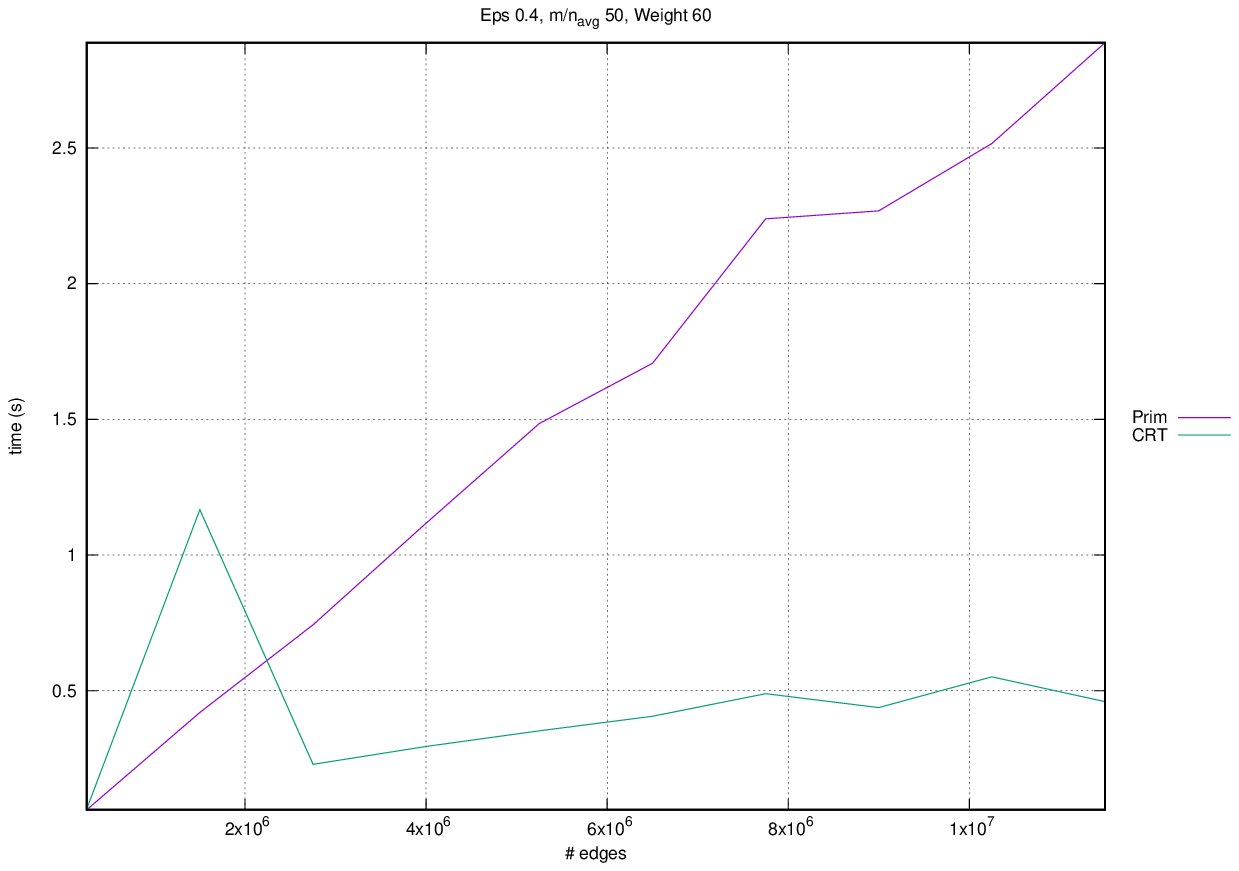}\label{U_04_50_60_time2}}
 \subfloat[][$w = 80$]
 {\includegraphics[width=.65\textwidth]{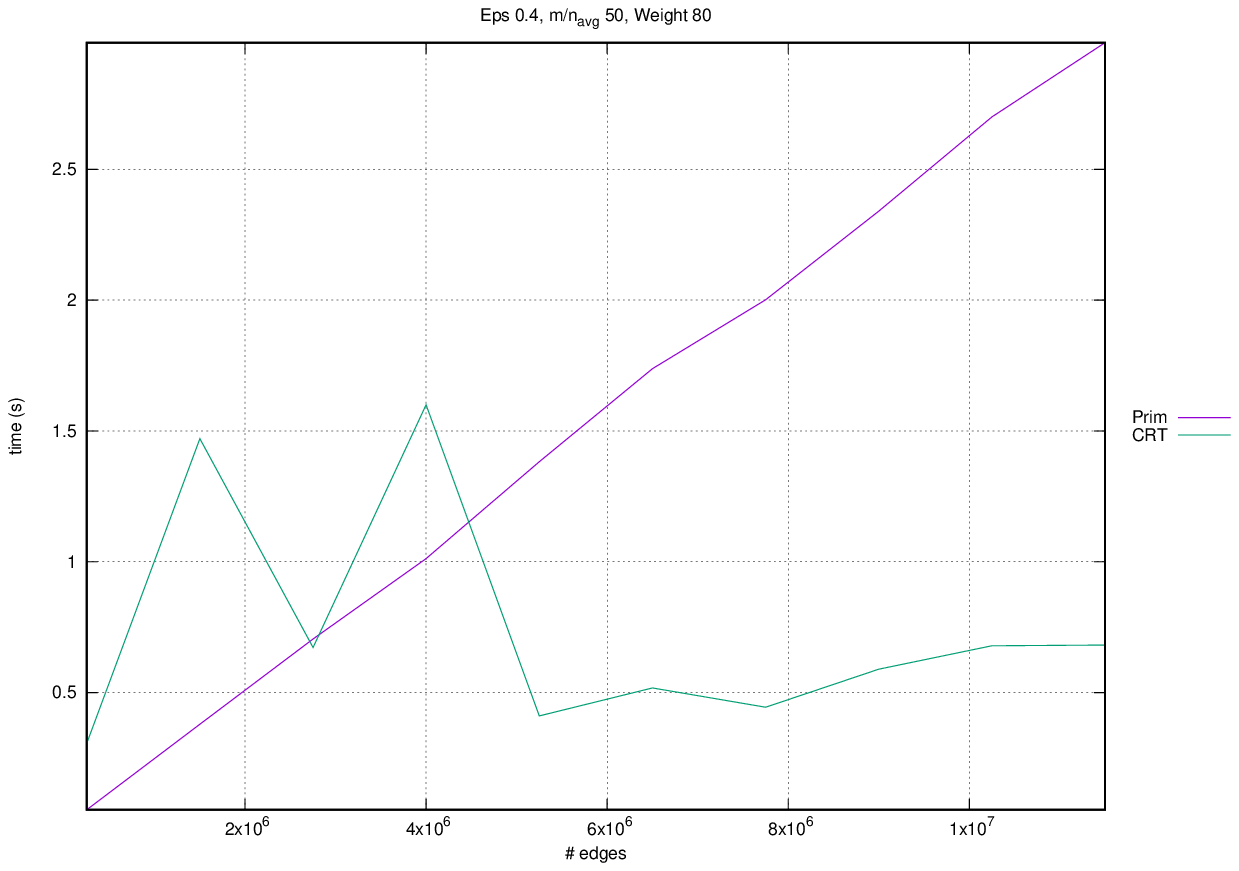}\label{U_04_50_80_time}}
 \caption{Behaviour for the increase of $w$. ({\sf Uniform model})}
 \label{w_increase_time}
\end{figure}

\begin{figure}[htbp]
 \centering
 \subfloat[][$w = 20$]
 {\includegraphics[width=.65\textwidth]{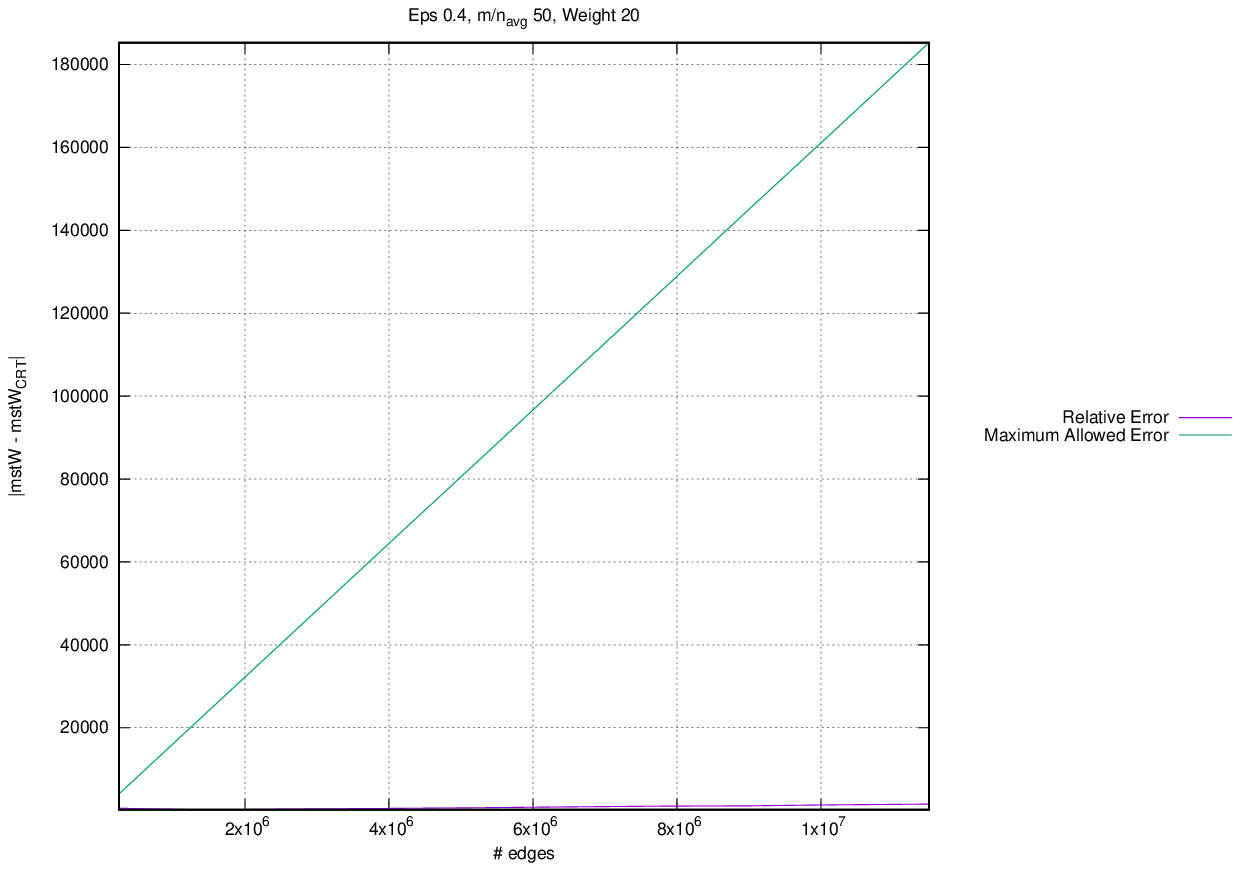}\label{U_04_50_20_rel}}
 \subfloat[][$w = 40$]
 {\includegraphics[width=.65\textwidth]{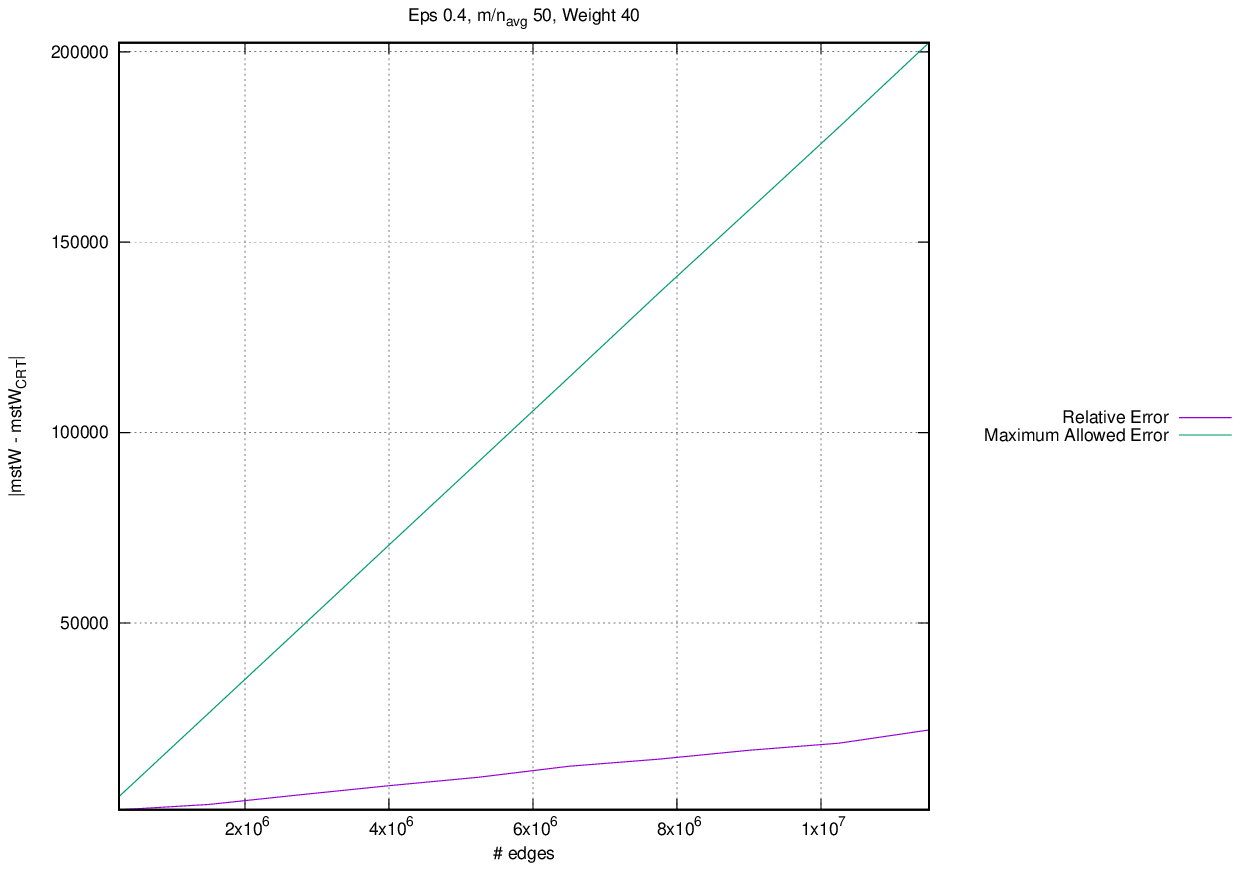}\label{U_04_50_40_rel}} \\
 \subfloat[][$w = 60$]
 {\includegraphics[width=.65\textwidth]{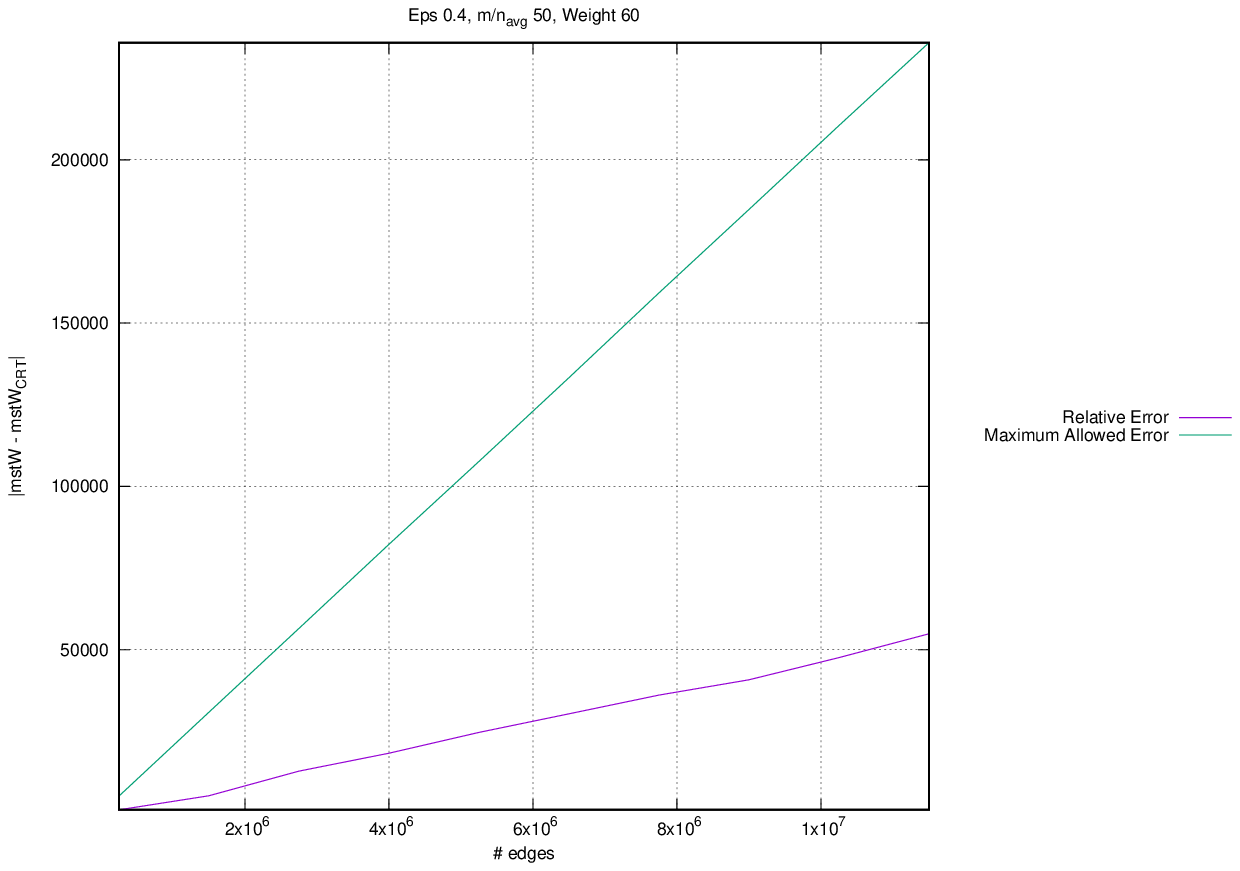}\label{U_04_50_60_rel2}}
 \subfloat[][$w = 80$]
 {\includegraphics[width=.65\textwidth]{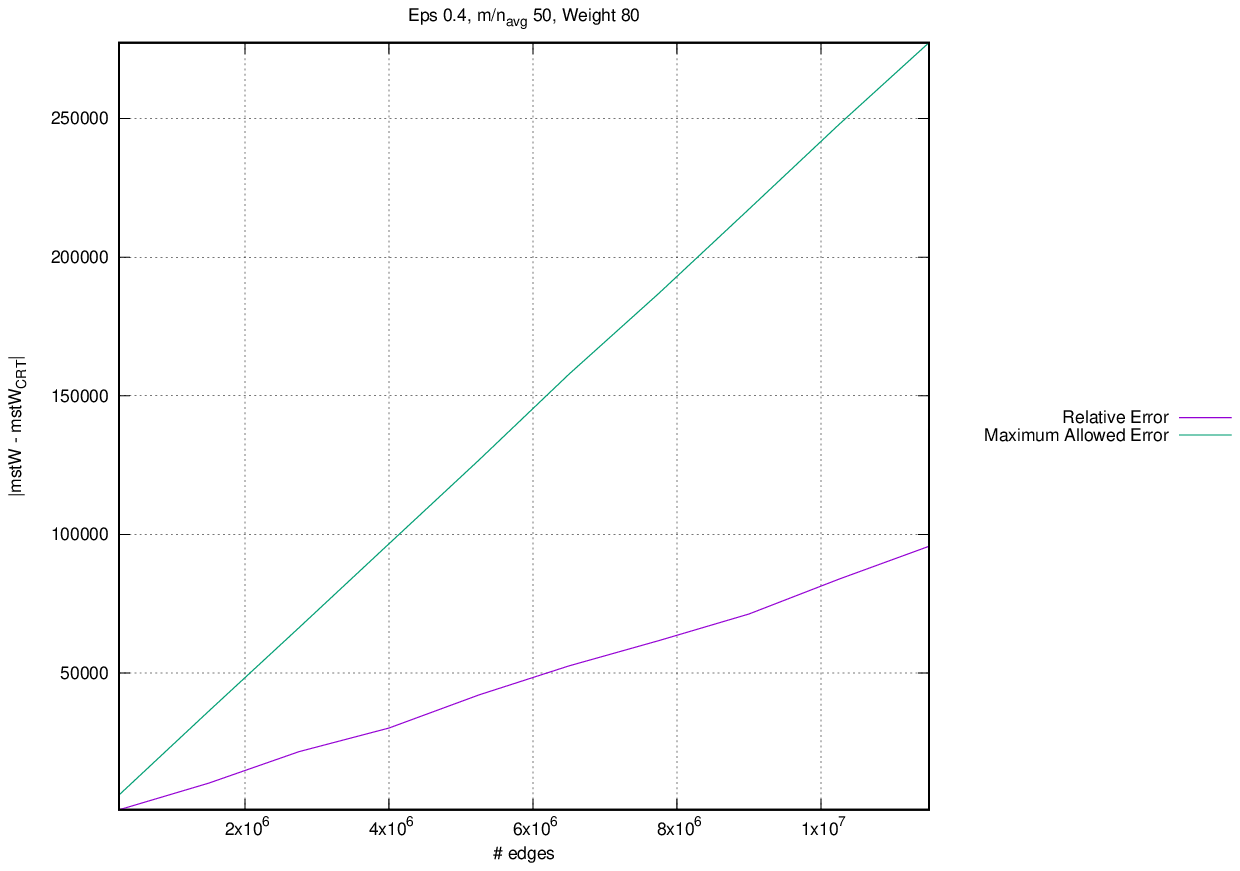}\label{U_04_50_80_rel}}
 \caption{Error behaviour for the increase of $w$. ({\sf Uniform model})}
 \label{w_increase_rel}
\end{figure}

Here we will manipulate the value of $w$, similarly to what we have already done with the average degree. Figures~\ref{w_increase_time} and~\ref{w_increase_rel} are hereafter proposed; this time the other fixed values are $\varepsilon = 0.4$, $d = 50$. Still this graphs has been build using a uniform model.

This time we see the error growing as $w$ increase. So we see here a \emph{direct} proportion of the execution time with the maximum weight, unlike the \emph{inverse} proportion it had with $d$. This is due to the fact that every iteration of the subroutine \texttt{approx-number-connected-components} that the reader can find in the original paper and remembered in pseudocode~\ref{alg}, adds a further approximation to the final result, because approximates the addend $\hat{c}$ that contributes to the total approximation, that is, the final error.

We see also that the dimension of the initial curve described here so far, grows proportionally to the worsening of the excution time's trend. This evidence also is observable in all the trends studied.

\subsection{Error tolerance $\varepsilon$}

\begin{figure}[htbp]
 \centering
 \subfloat[][$\varepsilon = 0.2$]
 {\includegraphics[width=.65\textwidth]{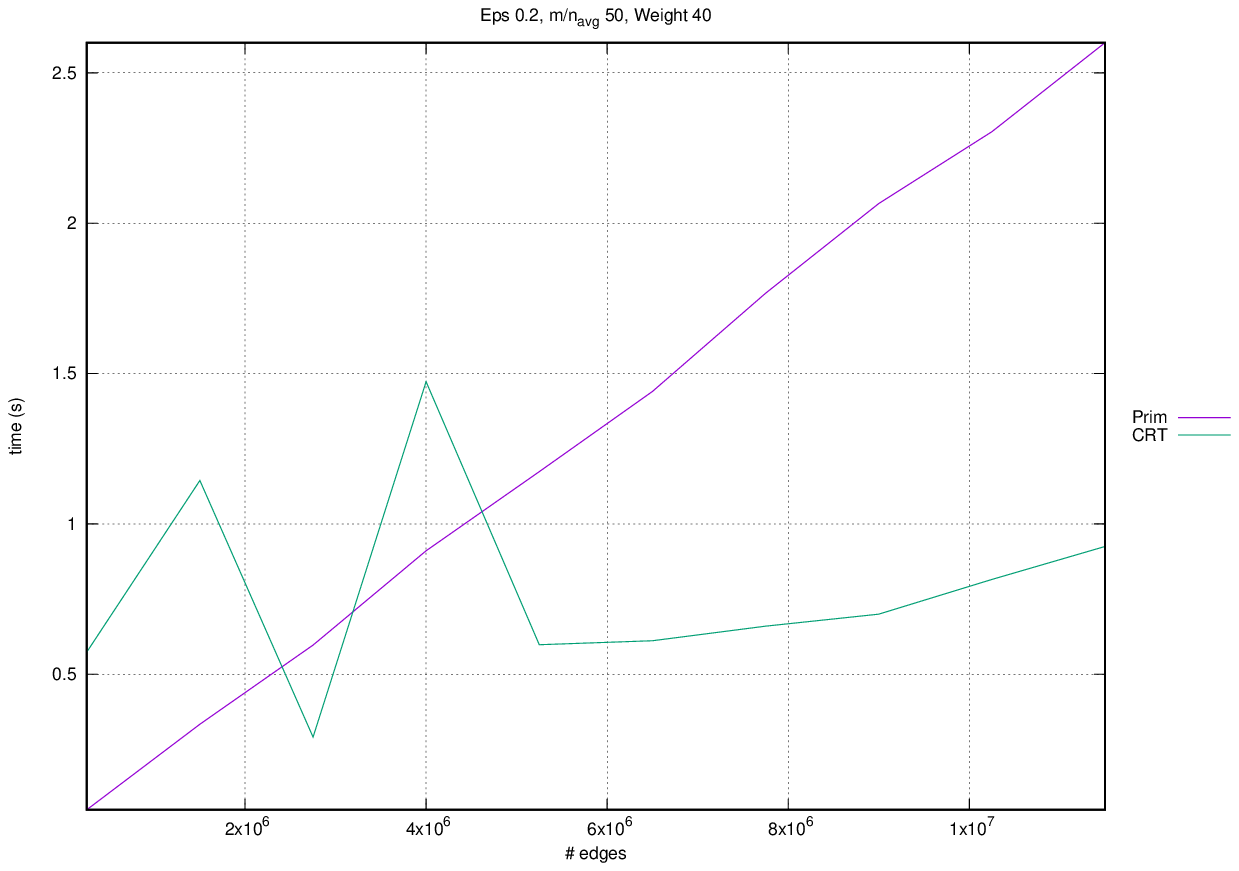}\label{U_02_50_40_time}}
 \subfloat[][$\varepsilon = 0.3$]
 {\includegraphics[width=.65\textwidth]{plots/uniform_03_50_40_time}\label{U_03_50_40_time2}} \\
 \subfloat[][$\varepsilon = 0.4$]
 {\includegraphics[width=.65\textwidth]{plots/uniform_04_50_40_time}\label{U_04_50_40_time2}}
 \subfloat[][$\varepsilon = 0.49999$]
 {\includegraphics[width=.65\textwidth]{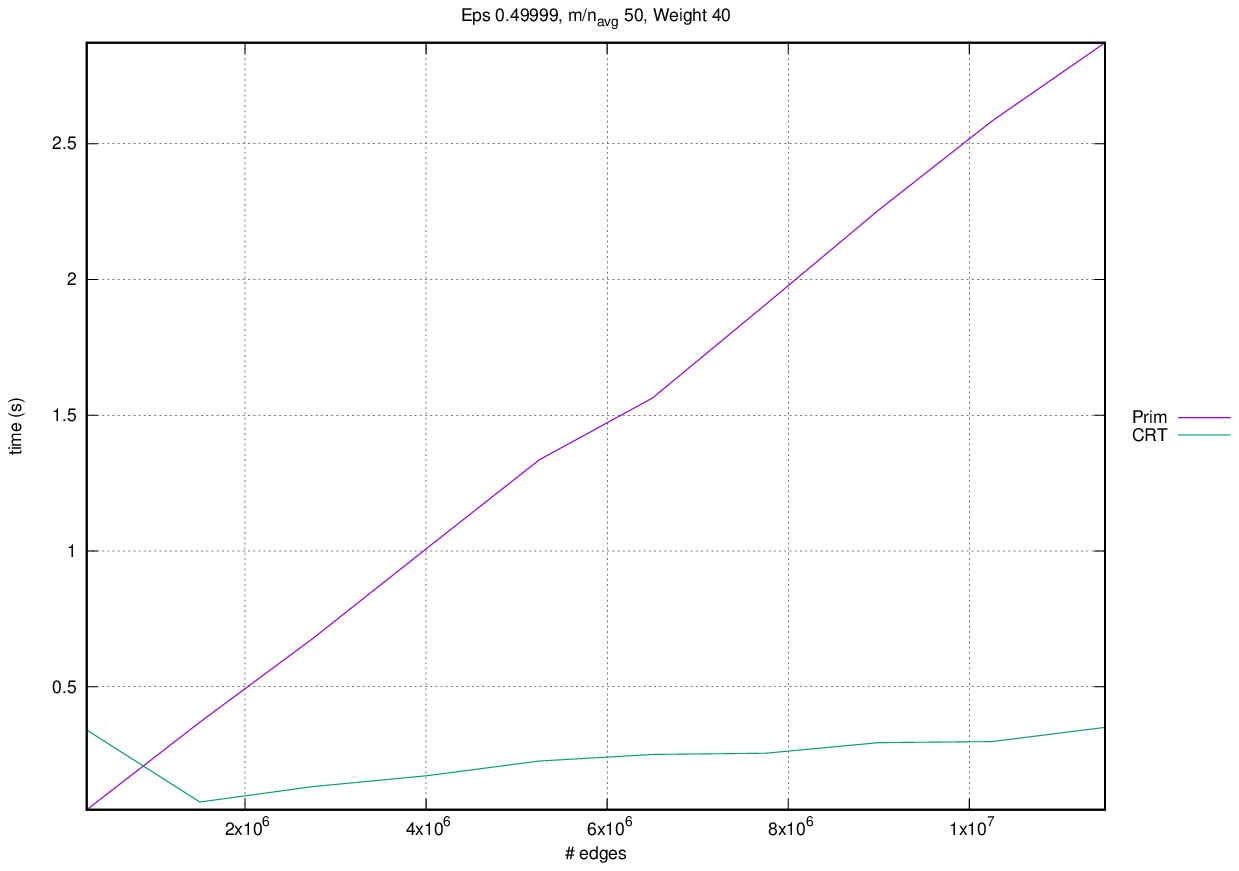}\label{U_049999_50_40_time}}
 \caption{Behaviour for the increase of $\varepsilon$. ({\sf Uniform model})}
 \label{e_increase_time}
\end{figure}

\begin{landscape}

\begin{figure}[htbp]
 \centering
 \subfloat[][$\varepsilon = 0.2$]
 {\includegraphics[width=.411\textwidth]{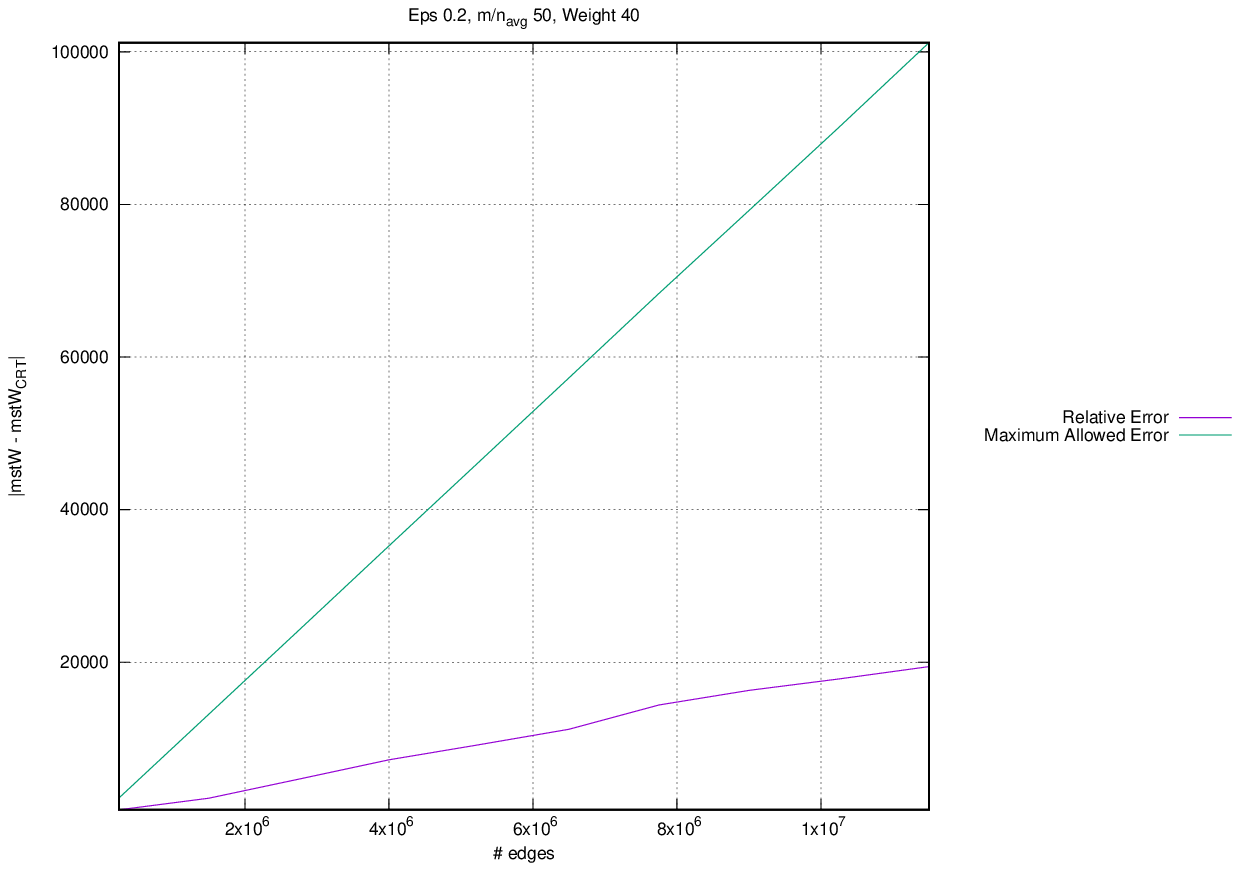}\label{U_02_50_40_rel}}
 \subfloat[][$\varepsilon = 0.3$]
 {\includegraphics[width=.411\textwidth]{plots/uniform_03_50_40_rel}\label{U_03_50_40_rel3}}
 \subfloat[][$\varepsilon = 0.4$]
 {\includegraphics[width=.411\textwidth]{plots/uniform_04_50_40_rel}\label{U_04_50_40_rel2}}
 \subfloat[][$\varepsilon = 0.49999$]
 {\includegraphics[width=.411\textwidth]{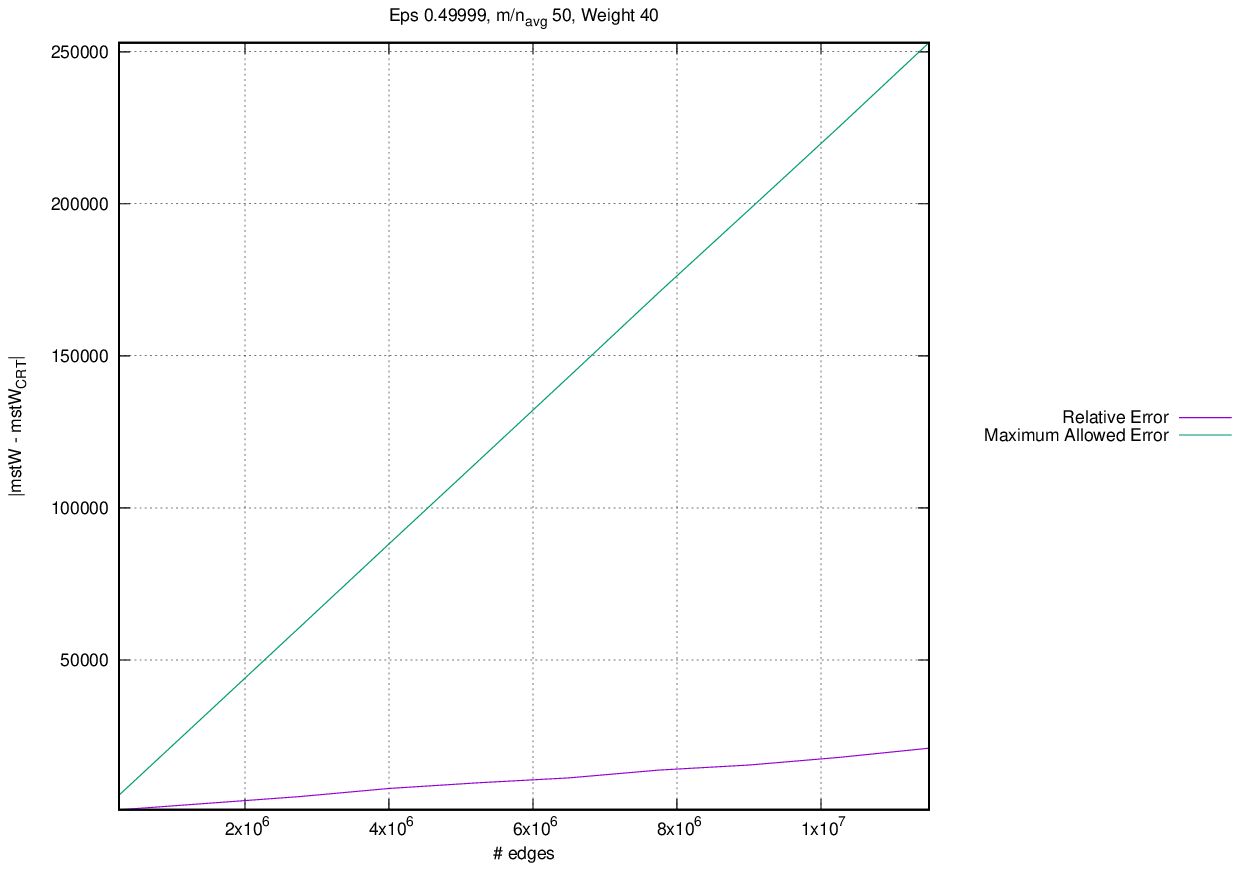}\label{U_049999_50_40_rel}} \\
  \subfloat[][$\varepsilon = 0.2$]
 {\includegraphics[width=.411\textwidth]{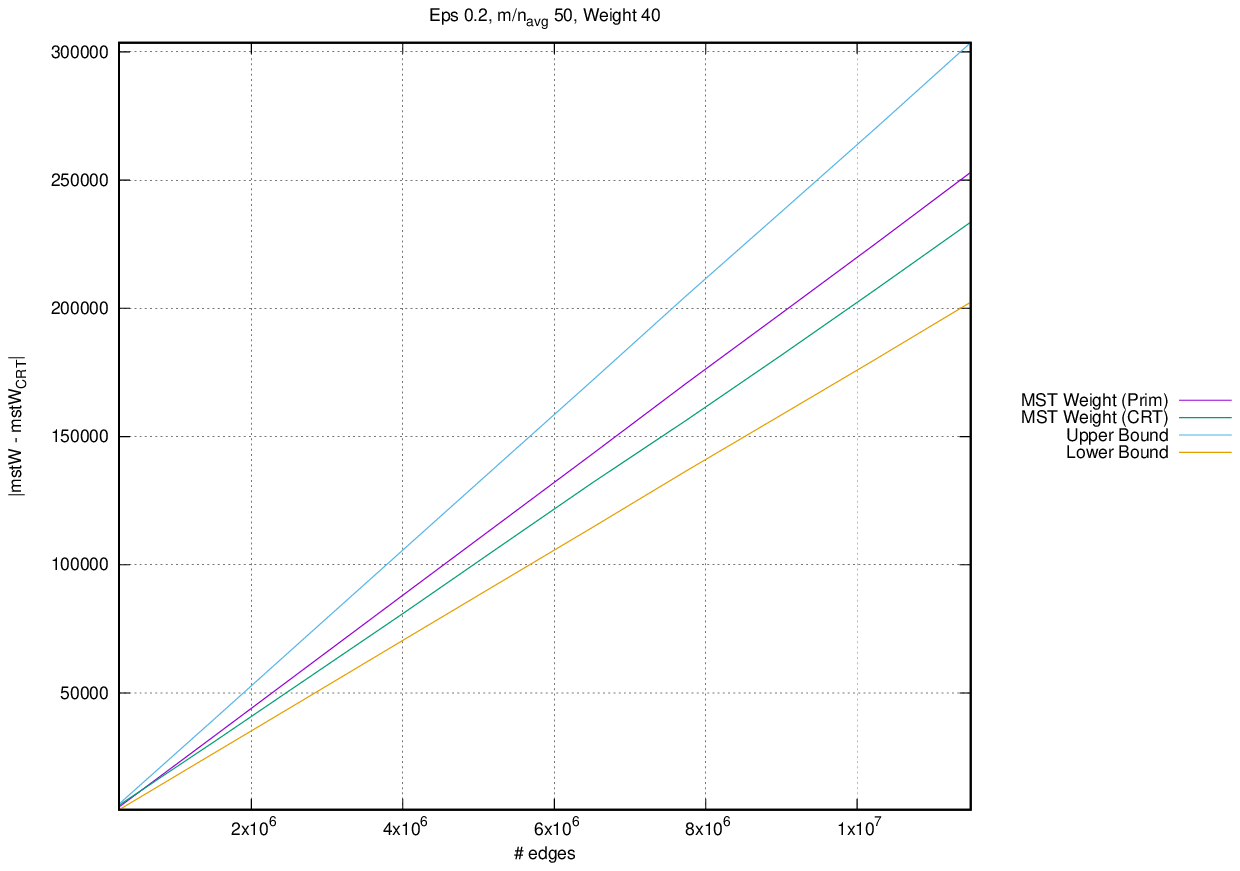}\label{U_02_50_40_abs}}
 \subfloat[][$\varepsilon = 0.3$]
 {\includegraphics[width=.411\textwidth]{plots/uniform_03_50_40_abs}\label{U_03_50_40_abs2}}
 \subfloat[][$\varepsilon = 0.4$]
 {\includegraphics[width=.411\textwidth]{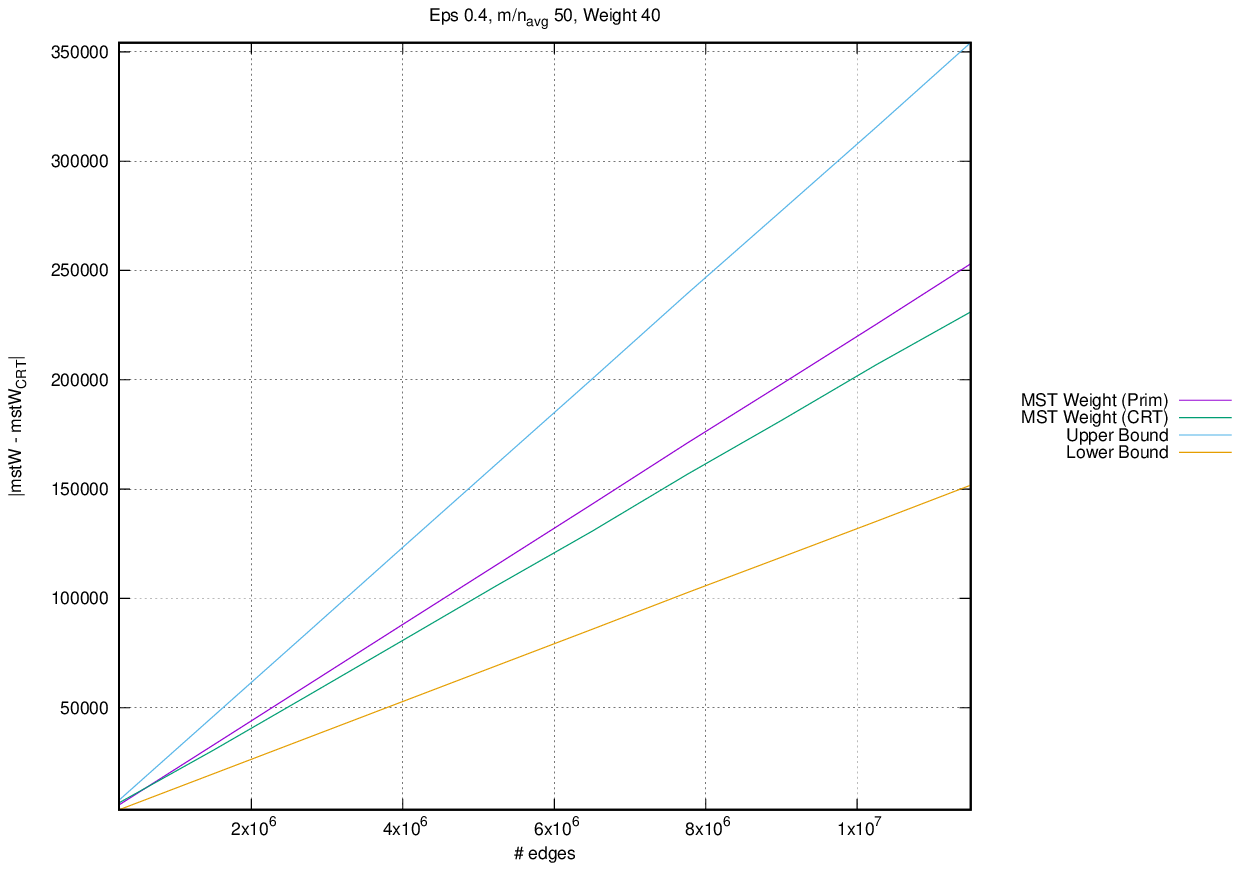}\label{U_04_50_40_abs}}
 \subfloat[][$\varepsilon = 0.49999$]
 {\includegraphics[width=.411\textwidth]{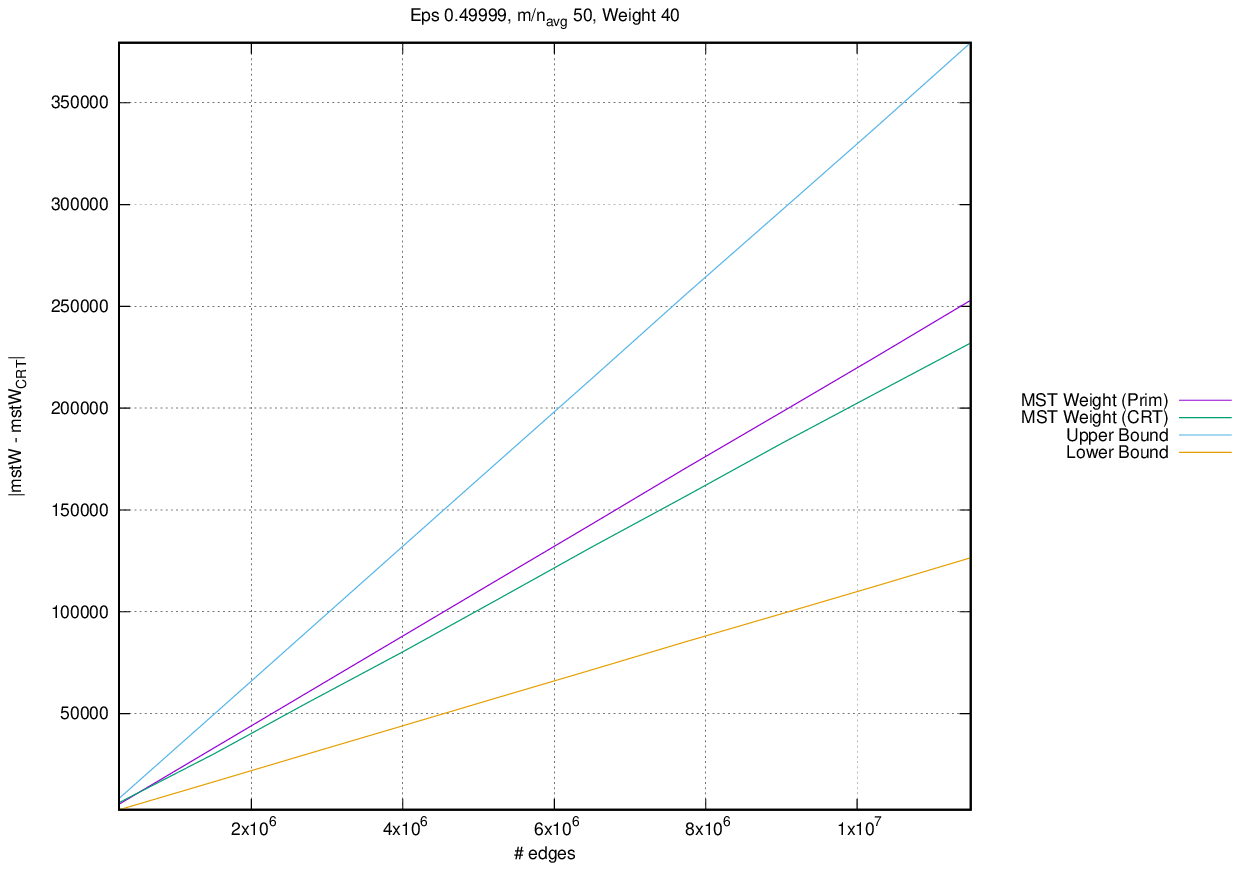}\label{U_049999_50_40_abs}} \\
 \caption{Absolute and relative value of the error in a {\sf Uniform} dataset, as $\varepsilon$ changes; fixed values are $\bar{d} = 50$, $w_\textup{max} = 40$.}
 \label{e_increase_rel}
\end{figure}

\end{landscape}

We will test our algorithm for the values of $\varepsilon = 0.2, 0.3, 0.4, 0.49999$ over uniform generated graphs. As already explained, no values below $0.2$ are investigated. We see in figures~\ref{e_increase_time} and~\ref{e_increase_rel} the trends.

As expected, we do note that the time trends tend to decrease as $\varepsilon$ increases, since we tolerate a higher error for the computed value, so the algorithm is less aggressive on computation and takes less time.

For the error trend instead we note an increase, still as expected; figures from~\ref{U_02_50_40_rel} to~\ref{U_049999_50_40_rel} has to be read with attention because every function graph has a different scale. The reader must not confuse the apparent lowering of the error since it is not the case. Looking at the other graphs about the absolute error, from~\ref{U_02_50_40_abs} to~\ref{U_049999_50_40_abs}, we see an expansion of the tolerance cone (light blue and ochre yellow lines) as increasing $\varepsilon$ means admitting an higher deviation from the correct MST weight value. Here too the different scaling must not confuse about the increasing trend of the error.

We see that the result is coherent with the theoretical model as the error increases with $\varepsilon$, but his variation is, after all, contained.

\subsection{Variations of graph model}

\begin{landscape}

\begin{figure}[htbp]
 \centering
 \subfloat[][uniform, time trend]
 {\includegraphics[width=.411\textwidth]{plots/uniform_04_50_60_time}\label{U_04_50_60_time}}
 \subfloat[][gaussian, time trend]
 {\includegraphics[width=.411\textwidth]{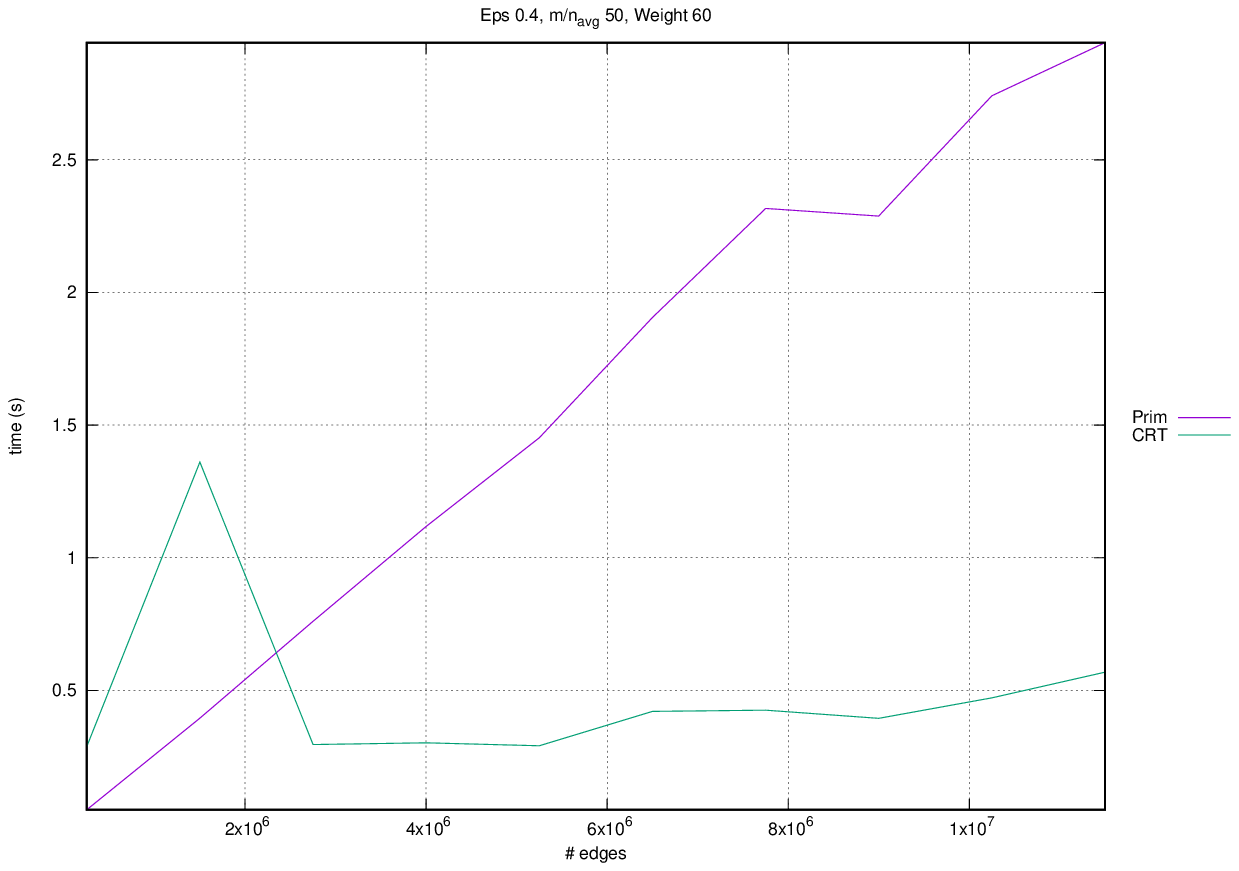}\label{G_04_50_60_time}}
 \subfloat[][small world, time trend]
 {\includegraphics[width=.411\textwidth]{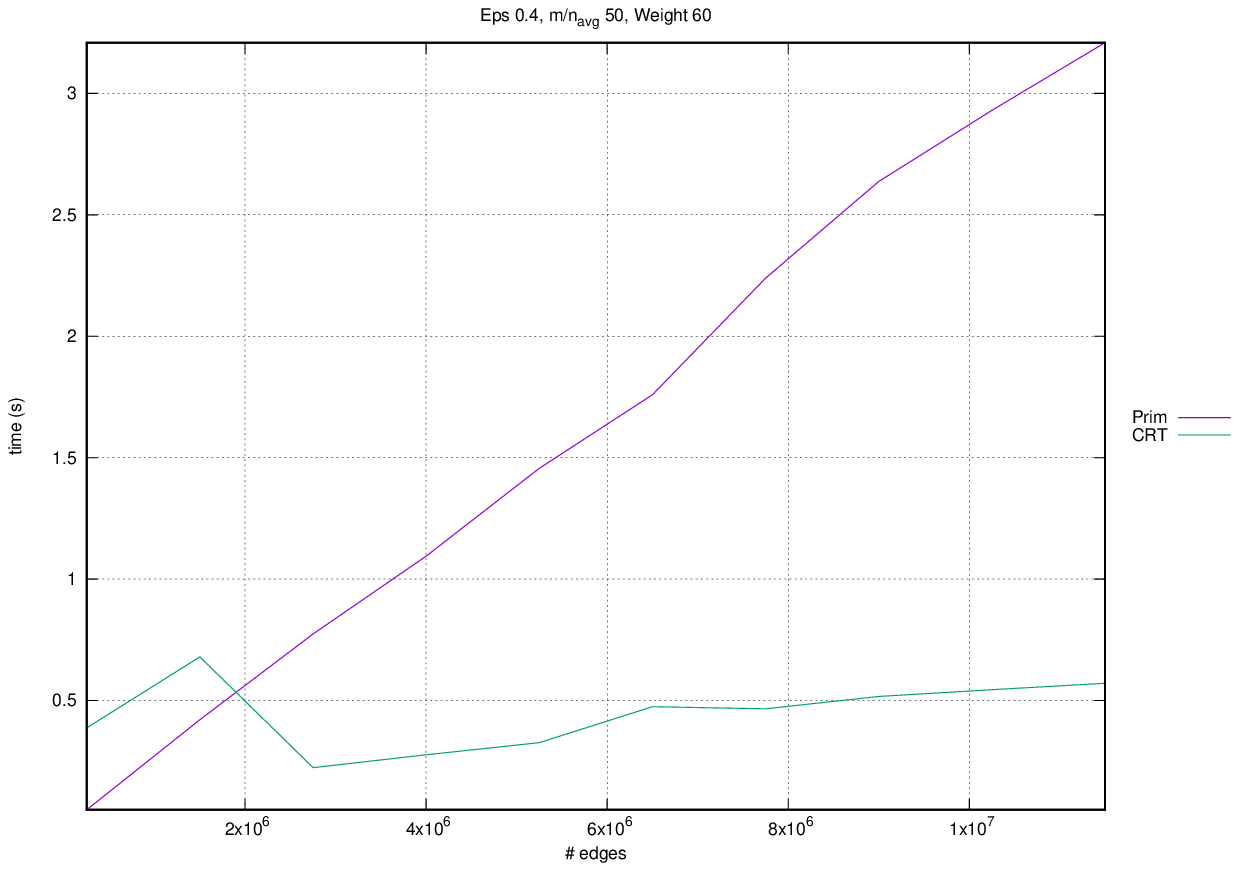}\label{SW_04_60_40_time}}
 \subfloat[][scale free, time trend]
 {\includegraphics[width=.411\textwidth]{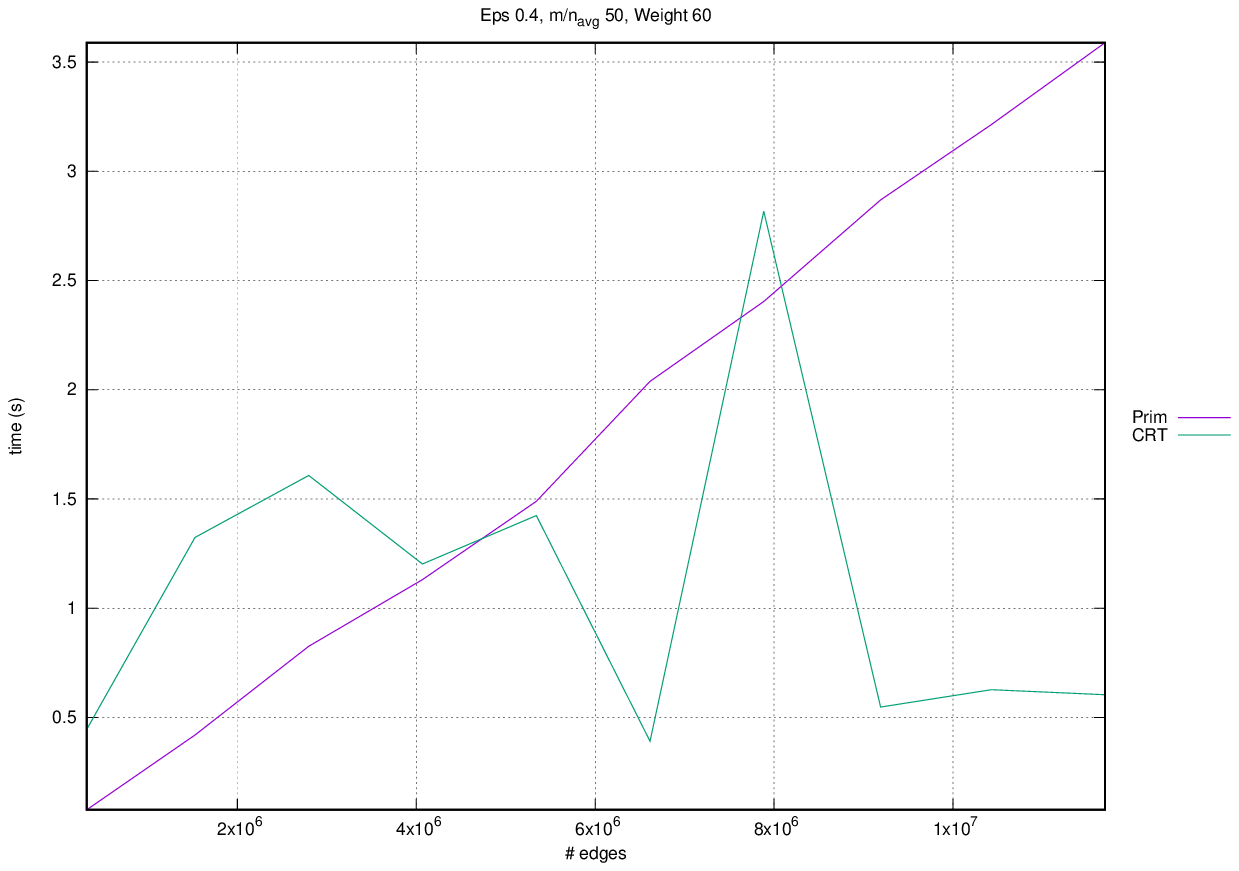}\label{SF_04_60_40_time}} \\
 \subfloat[][uniform, relative error]
 {\includegraphics[width=.411\textwidth]{plots/uniform_04_50_60_rel}\label{U_04_50_60_rel}}
 \subfloat[][gaussian, relative error]
 {\includegraphics[width=.411\textwidth]{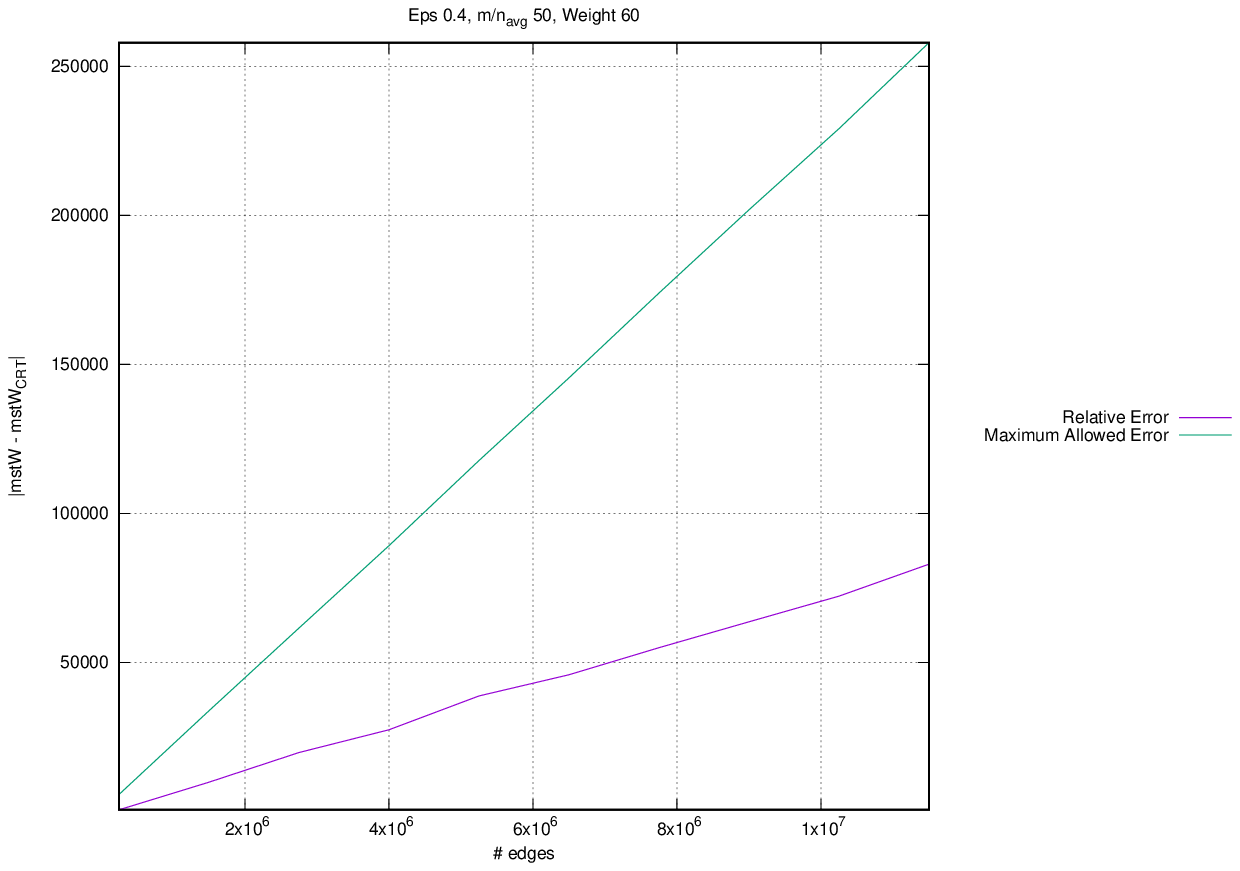}\label{G_04_50_60_rel}}
 \subfloat[][small world, relative error]
 {\includegraphics[width=.411\textwidth]{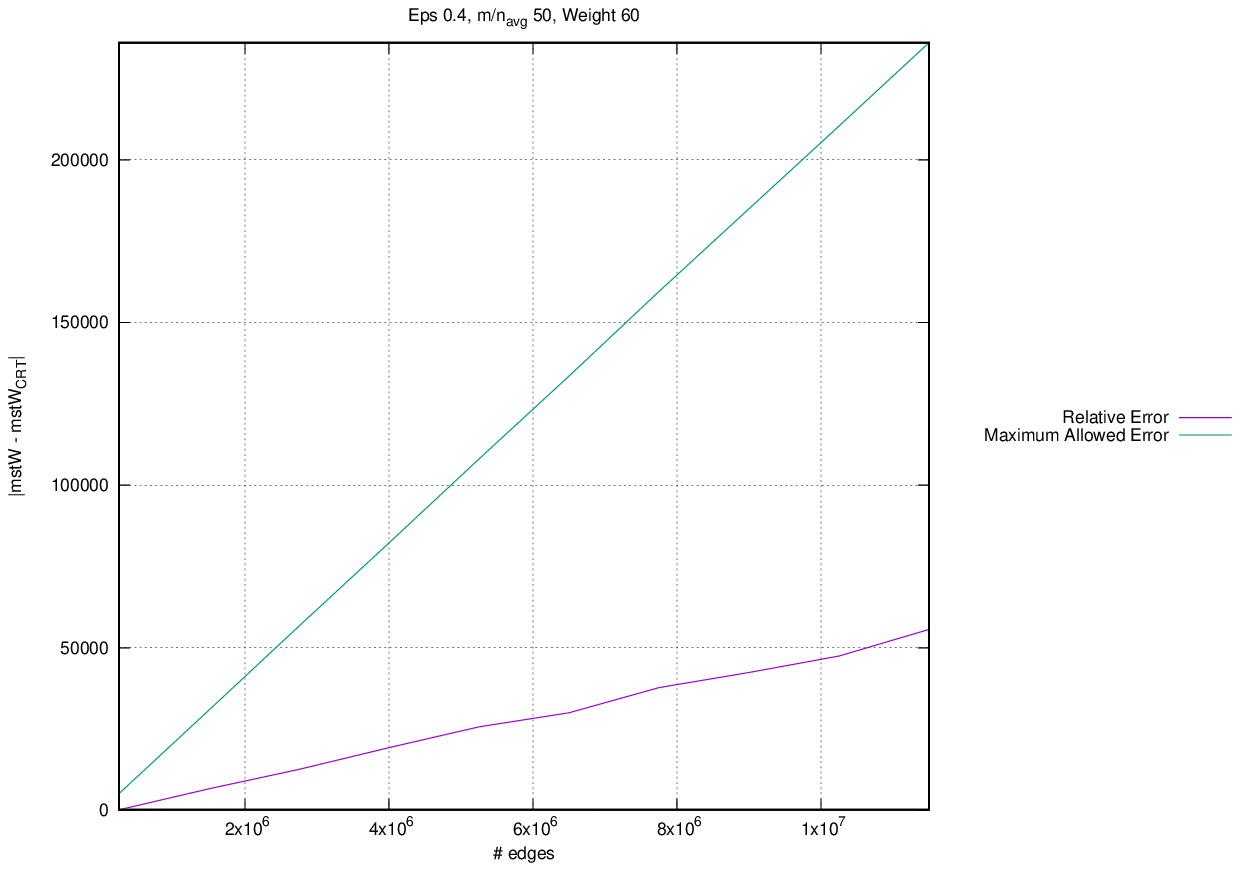}\label{SW_04_50_60_rel}}
 \subfloat[][scale free, relative error]
 {\includegraphics[width=.411\textwidth]{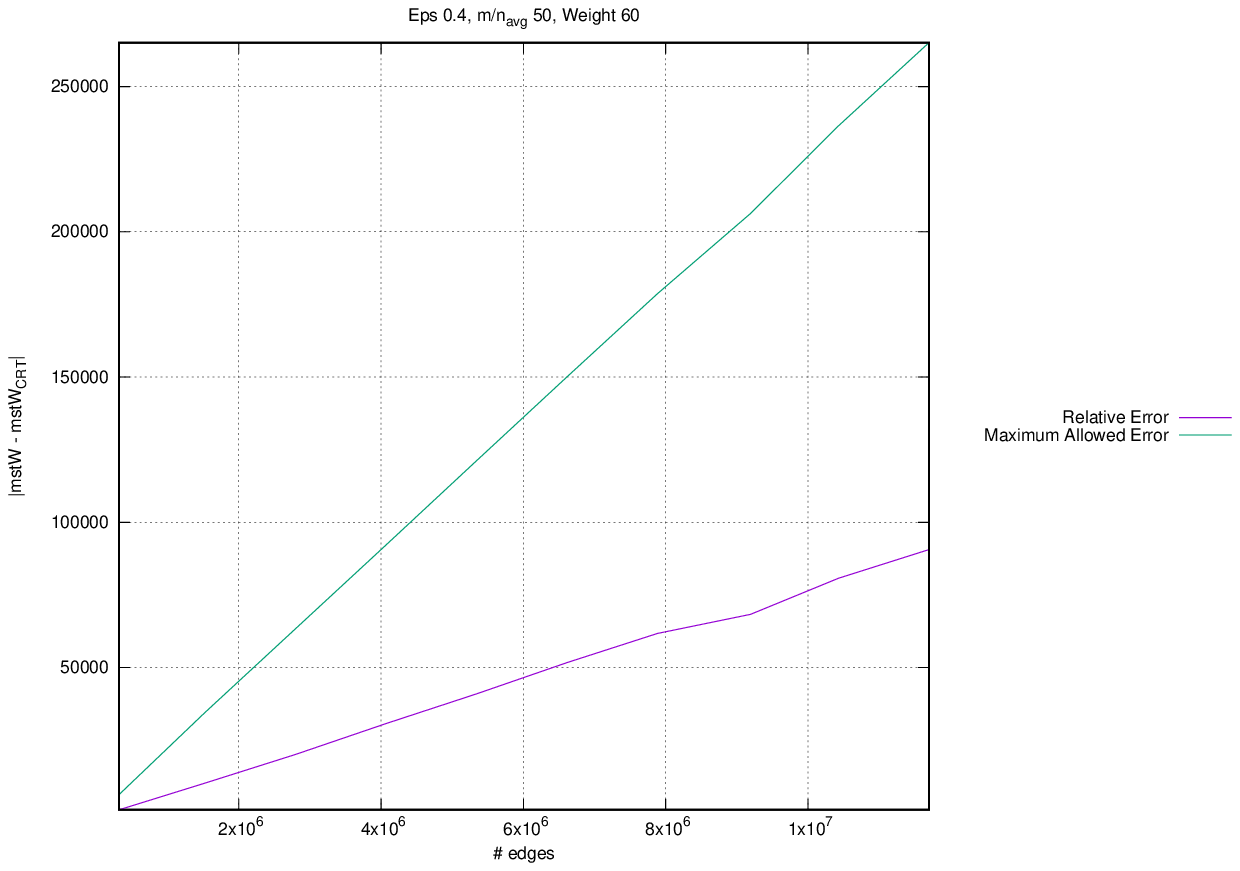}\label{SF_04_50_60_rel}}
 \caption{Different behaviours for different graph models; fixed values are $\varepsilon = 0.4$, $\bar{d} = 50$, $w_\textup{max} = 60$.}
 \label{model_variation}
\end{figure}

\end{landscape}

\begin{figure}[htbp]
 \centering
 \subfloat[][{\sf uniform}, $\bar{d} = 100$]
 {\includegraphics[width=.65\textwidth]{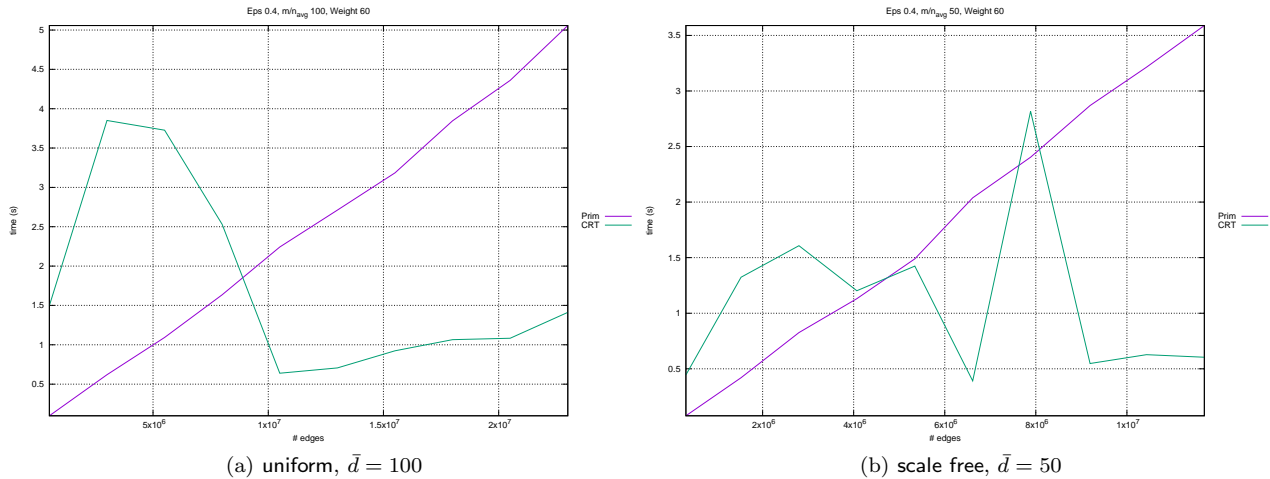}\label{U_04_100_60_time}}
 \subfloat[][{\sf scale free}, $\bar{d} = 50$]
 {\includegraphics[width=.65\textwidth]{plots/scalefree_04_50_60_time}\label{SF_04_50_60_time2}}
 \caption{A more correct comparison between the general case and scale-free graphs ($\varepsilon = 0.4$, $w_\textup{max} = 60$.)}
 \label{model_variation_part}
\end{figure}

As a last comparison, instead of varying one by one the crucial parameters and see the time and error trends, we fix all of them and try to change the model the graph belongs to. Figures~\vref{model_variation} show those results. We see here that both uniform and small-world models keep the same trend for the error, but the small-world one behaves slightly better on small instances. On the contrary, the gaussian show the same time trend respect to the uniform case, but his error has a higher growth curve. The scale-free model seems to be the worse case both regarding time and error trend, but we might remember, as observed earlier at~\vref{ddd}, that a real term of comparison requires to considerate a scale factor of $2$ as done in figure~\vref{model_variation_part}. We see in fact that, compared to the uniform case, the scale-free model has even a better behaviour, looking carefully at the function graph scale: but after all, both of them show sublinear complexity for instances of more than $10^7$ edges, so have both the same transients as the same steady state trend.

Another thing we note is that a bad trend in execution times are always bond to an explosion of the error, as we can see in the charts so far. This means that using more time to compute the value doesn't mean it will be nearer to the correct one.

\section{A specific case of study}\label{gbb}

\begin{figure}[htbp]
 \centering
 \subfloat[][Heavy case of \textsf{uniform}, $\varepsilon = 0.3$, $d \simeq 200$, $w = 80$]
 {\includegraphics[width=.5\textwidth]{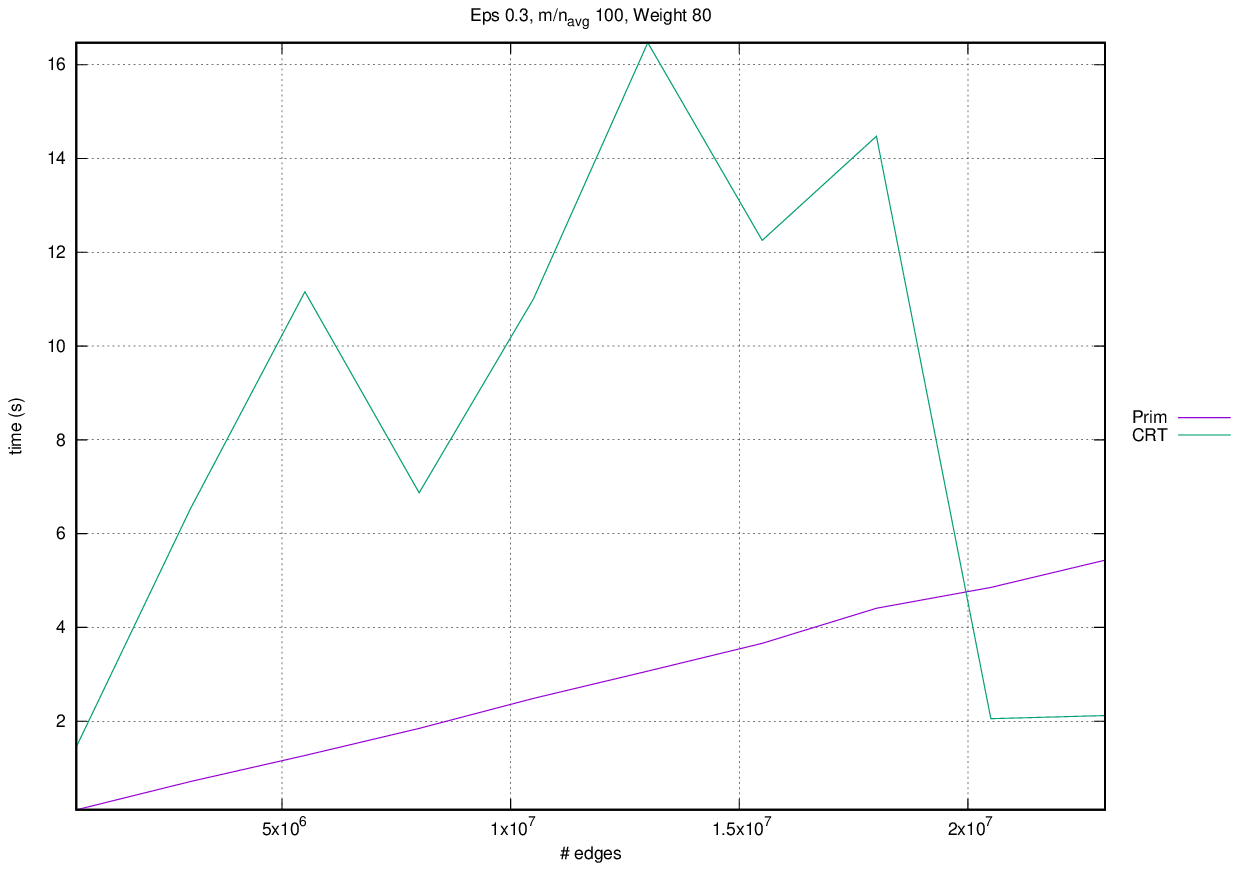}\label{U_03_100_80_time}}
 \subfloat[][Long run, up to 50 millions edges]
 {\includegraphics[width=.7\textwidth]{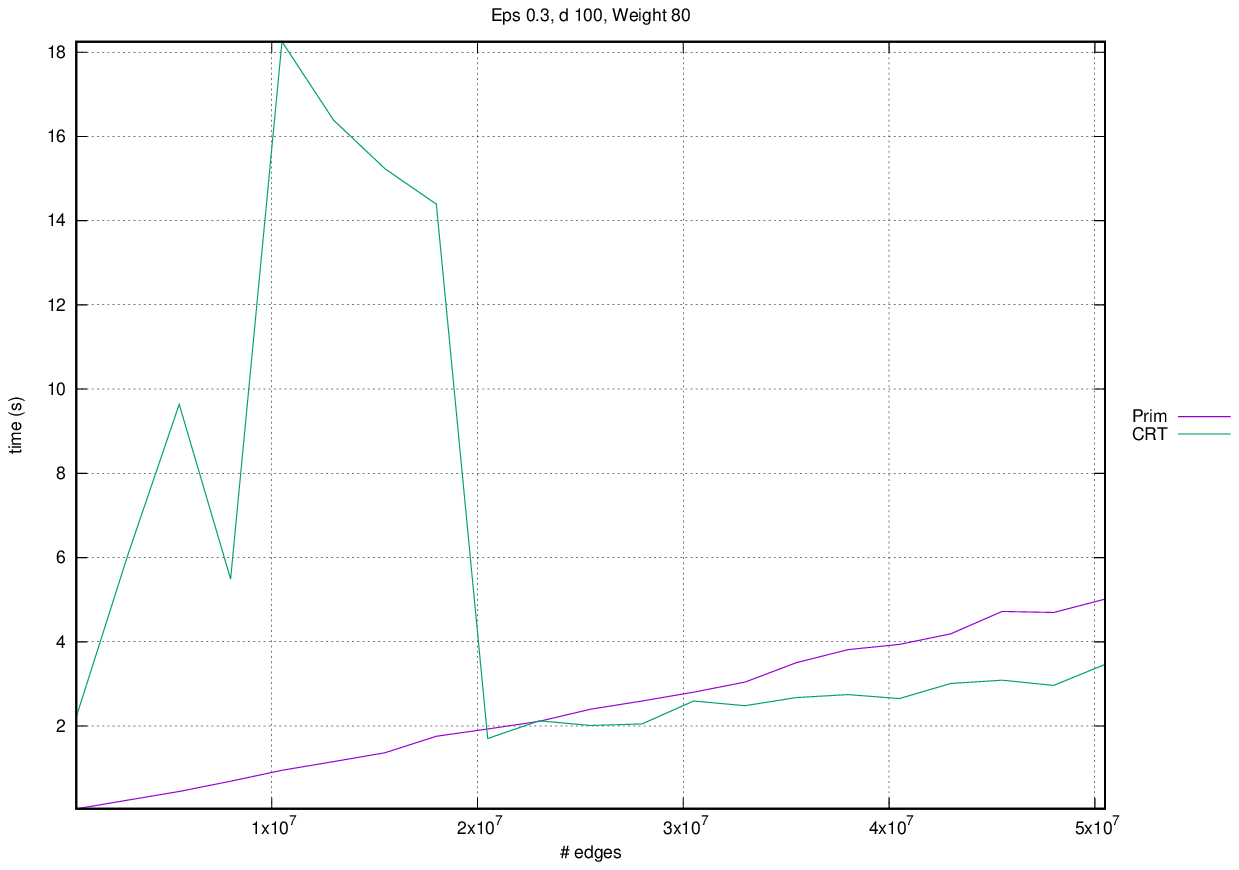}\label{bg_time}} \\
 \subfloat[][Long run, error trend]
 {\includegraphics[width=.75\textwidth]{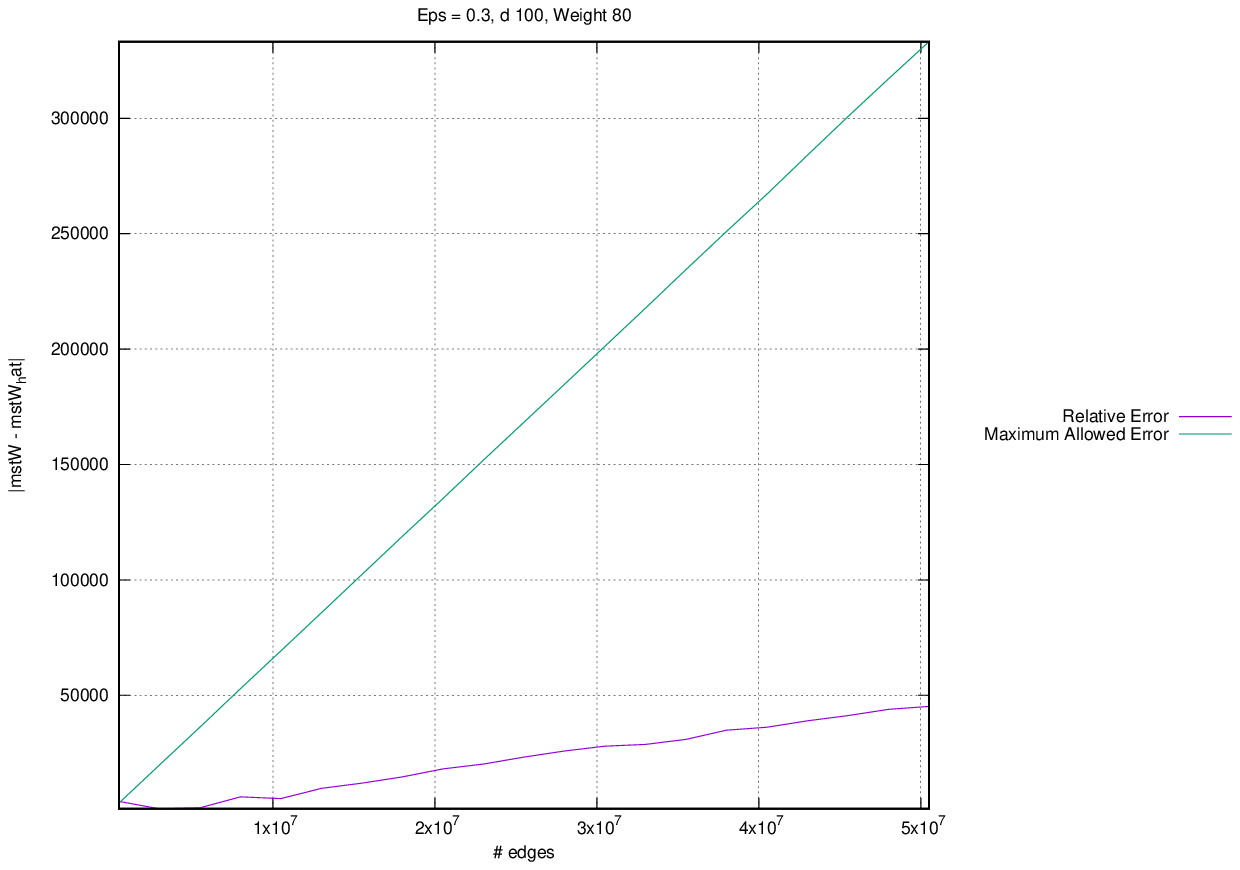}\label{bg_err}}
 \caption{Long run results; figure (a) shows the original case of study, while (b) his extension.}
 \label{long_run}
\end{figure}

At this point, all of our graphs show a bad initial curve every time we burden the input instance; in a specific case of study this anomalous curve was so persistent that all the function graph was more than linear; we decided to deepen this special case of study, performing a longer run to see the tail of this particular trend. The results are reported on figure~\vref{long_run}.

We can see that the original case of study hinted a sublinear trend beyond 20 millions edges instances, but we considered to investigate further, and figure~\ref{bg_time} confirms the sublinearity. We have a good behaviour on the error trend (figure~\ref{bg_err}).

\section{Conclusions and final thoughts}

As expected, a probabilistic algorithm like the CRT allows us to compute an approximation of the Minimum Spanning Tree weight in sublinear time on the number of edges, under certain conditions.
Tunable parameters, that depends on $\varepsilon$, allows us to perform either a better or a worse approximation, implying respectively a very slow and a very fast computation. The choice of a small value of $\varepsilon$ can lead to terrible running times, and for these values it does not make sense to compare the CRT algorithm with any other deterministic algorithm.\\
For other $\varepsilon$ values, instead, we prove the good performances of the CRT. The reader can easily view the better performances of CRT algorithm versus Prim algorithm or Kruskal algorithm watching the line charts in the previous section of this paper.

More in general, we see that execution time and error depend on the number of BFSes successfully completed during the computation The more BFSes are completed, the more information the algorithm has to predict a correct value for the MST weight, but on the other hand, completing the BFSes takes time. If instead we have a lot of interrupted BFSes, we waste a large amount of time without gathering information, hence resulting in both high execution time and error.

We considered so far different theoretical graph models, and we conclude that a high \emph{clustering coefficient} tends to increase the probability to have interrupted BFSes. This because one of the reasons the algorithm has to interrupt the search is having encountered a hub, i.e. a vertex with a high degree. We saw in fact that when changing the model there is a slight perturbation of the trend, although it remains sublinear. The key concept is the ``distance'' from the hubs of the graph of the root vertex from which our BFS starts.

Generally we also concluded that increasing the average degree let our algorithm gather more information, because there is a growing of the probability to visit the generic node $u$ of our graph. On the other side, increasing the maximum weight correspond to an increase in the number of iterations our algorithm does, that leads to summing more intermediate approximated results that imply a higher final approximation.

\subsection{Parallel implementation}
We observe that the CRT algorithm lends itself very well to a parallel implementation. Indeed the majority of the algorithm's code is organized into independent sections, and in most cases they don't need to comunicate to each other. We also observe that three levels of paralellism can be achieved within the code. In the first level we parallelize each of the $w$ independent calls to \texttt{approx-number-connected-components}, as depicted in pseudocode~\ref{alg}, below; every of this calls internally performs $r$ independent BFSes from $r$ different roots, that could in turn run in parallel, achieving a second level of parallelism. Moreover, considering that in the academic world there already exist different parallel implementations of the BFS algorithm, we can use one of them to perform an additional third level of paralellism.

\begin{algorithm}
   \caption{First level parallelism for the CRT algorithm}
   \label{alg}
    \begin{algorithmic}[*]
     \Function{approx-MST-weight}{$G$, $\varepsilon$}\Comment{$G$ - input graph, $\varepsilon$ - error tolerance}
	
	\State $d^* \gets $ \Call{approx-avg-degree}{$\varepsilon$}\Comment{sequential, runs in $O(d/\varepsilon)$ as shown in~\cite{crt}}
	\State $\hat{c} \gets 0$
	
	\For{$i = 1, \dots , i = w$}
	    \Comment{{\small \hlgray{this $w-1$ calls can be run in $w-1$ parallel threads}}}
	    \State $\hat{c} \mathrel{+}=$ \Call{approx-number-connected-components}{$G^{(i)}$, $\varepsilon$, $d^*$}
	\EndFor
	
	\State \Return $\hat{v} \gets n - w + \hat{c}$
     \EndFunction
    \end{algorithmic}
\end{algorithm}

To make it even more simpler, the number of different flows is known a priori, so a static pre-instantiation and a smart scheduling of the threads can be performed. At the first and second levels a \emph{master-slave} model can be used in a \emph{fork-join} structure, while in the third level a shared variable is needed between the different BFSes. As a final remark, parallelizing the BFSes could have too much overhead given that the algorithm is optimized to run them very fast and to stop the ones that seem to cost too much.

\section{Future prospects}\label{rtn}

During an e-mail exchange with one of the original authors of \cite{crt}, Dr. Ronitt Rubinfeld, another topic of discussion and study has emerged, about the distribution of the weight on the edges.

\begin{figure}[htbp]
 \centering
 \subfloat[][Assuming a uniform distribution]
 {\includegraphics[width=.525\textwidth]{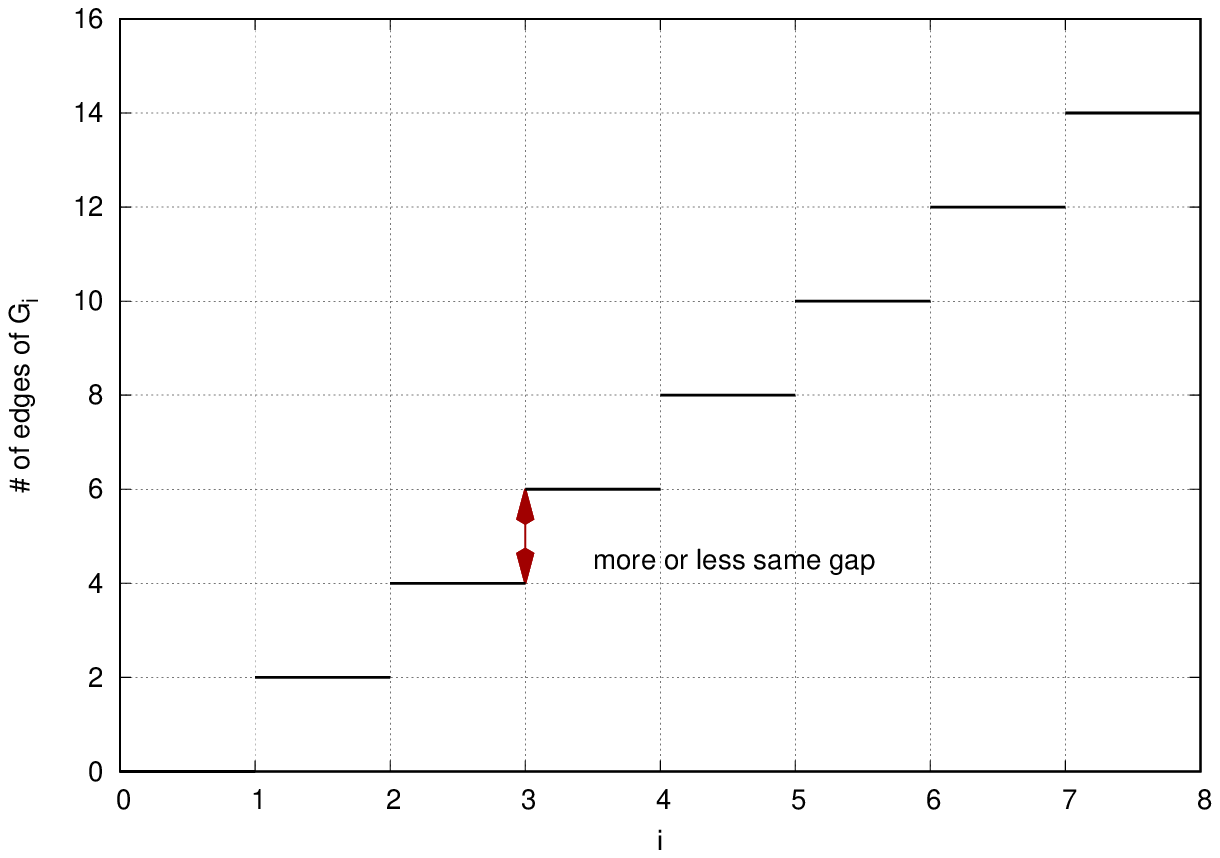}\label{same_gap}}
 \subfloat[][Assuming a generic distribution]
 {\includegraphics[width=.525\textwidth]{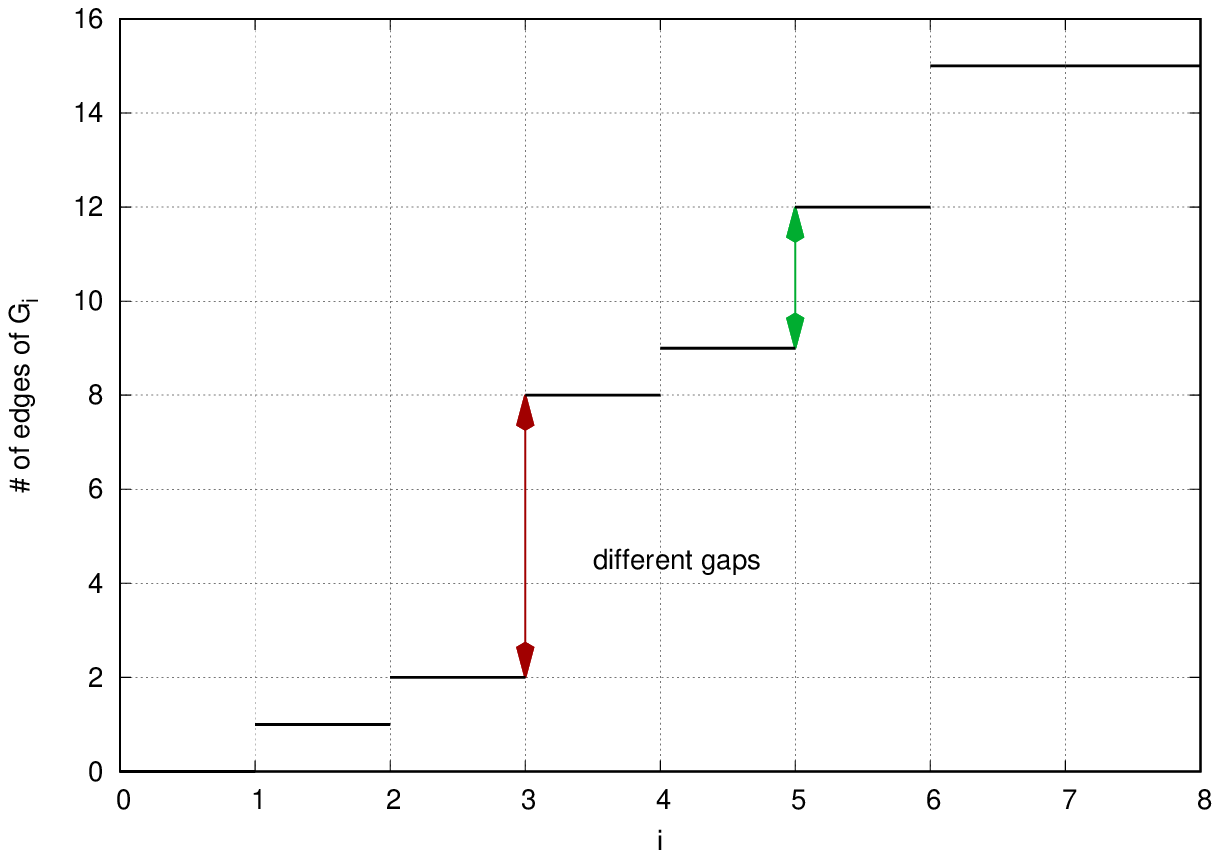}\label{diff_gap}}
 \caption{Expected behaviour with different laws for the distribution of weights.}
 \label{gaps}
\end{figure}

In our code, the generation of random graphs only assume a uniform distribution of the weights on the edges, i.e. a given edge can have an integer weight $k \in [1, w]$ with probability $\frac{1}{w}$. That implies a linear growth of the dimension of $E(G_i), \forall i \in [1, w]$, namely the set of $G_i$'s edges; this is well depicted in figure~\ref{same_gap}, where, as $i$ grows, the size of $E(G_i)$ increases at each step of a quantity ``near'' $\frac{ \left| E(G) \right| }{w}$, and it is more true as $\left| E(G) \right|$ is big for the law of large numbers. On the other side, having a generic law of distribution for the edges weight implies having a different behavior as depicted on~\ref{diff_gap}.

This difference could be of interest because it means that the input of different subsequent iterations of \texttt{approx-number-connected-components} will have a non regularly increasing size, or even the same size for some calls; it can be easily shown indeed that the function in~\ref{gaps} is nondecreasing. Since the cost of those calls finally determines the overall cost, we might argue that this could lead to a minor difference, but at the same time think that only with another set of tests we could conclude something relevant about this observation.


\end{document}